\documentclass[11pt]{iopart}
\usepackage{iopams}
\usepackage{epsfig}
\usepackage{amssymb}
\usepackage{color}
\usepackage{graphicx}
\usepackage{epstopdf}
\usepackage{float}
\usepackage{color}
\usepackage[margin=1.6cm]{geometry}
\usepackage[breaklinks=true,colorlinks=true,linkcolor=blue,urlcolor=blue,citecolor=red]{hyperref}
\begin{document}
\title[]{Nuclear shape phase transition within a conjonction of $\gamma$-rigid and $\gamma$-stable collective behaviours in  deformation dependent mass formalism }
\author{M. Chabab$^*$, A. El Batoul, A. Lahbas  and M. Oulne}
\address{High Energy Physics and Astrophysics Laboratory, Department of Physics, Faculty of Sciences Semlalia, Cadi Ayyad University P.O.B 2390, Marrakesh 40000, Morocco.}
\eads{\mailto{\textcolor[rgb]{0,0.35,1.0}{mchabab@uca.ma}($^*$Corresponding author)}, \mailto{\textcolor[rgb]{0,0.35,1.0}{elbatoul.abdelwahed@edu.uca.ma}},
	\mailto{\textcolor[rgb]{0,0.35,1.0}{alaaeddine.lahbas@edu.uca.ma}},
	 \mailto{\textcolor[rgb]{0,0.35,1.0}{oulne@uca.ma}}}
\begin{abstract}
	In this paper, we present  a theoretical study of a conjonction of $\gamma$-rigid and $\gamma$-stable collective motions in  critical point symmetries of the phase transitions  from spherical to deformed shapes of nuclei using exactly separable version of the Bohr Hamiltonian with deformation-dependent mass term. The deformation-dependent mass is applied simultaneously to  $\gamma$-rigid  and $\gamma$-stable parts of this famous collective Hamiltonian. Moreover, the $\beta$ part of the problem is described by means of  Davidson potential, while the  $\gamma$-angular part  corresponding to axially symmetric shapes  is treated by a Harmonic Osillator potential. The energy eigenvalues and normalized eigenfunctions of the problem  are obtained in compact forms by making use of  the asymptotic iteration method. The combined effect of the deformation-dependent mass and  rigidity as well as harmonic oscillator stiffness parameters on the energy spectrum and wave functions is duly investigated. Also, the electric quadrupole transition ratios and energy sprectrum  of some  $\gamma$-stable and prolate nuclei are calculated and compared with the experimental data as well as with other theoretical models.
\end{abstract}
\vspace{2pc}
\section{Introduction} 
\label{s1}
The study of shape phase transitions in nuclei, within the Bohr-Mottelson model \cite{b1,b2}, has known a particular interest during the last decade. Such an interest has grown even more with the occurrence of critical point symmetries (CPS) which rather are an experimental evidence. Such symmetries like, for example, $E(5)$ \cite{b3} corresponding to the second order phase transition between spherical  and $\gamma$-unstable nuclei and the $X(5)$ \cite{b4} symmetry which is designed to the first order phase transition between vibrational $U(5)$ and axially symmetric prolate $SU(3)$ nuclei, have motivated the search for new suitable  solutions of the Bohr Hamiltonian \cite{b1,b2} beyond the infinite square well potential originally used in both mentioned above critical points for the $\beta$ collective motion. Regarding the $\gamma$- vibrations, the corresponding potential was assumed to be independent of the $\gamma$-variable in the $E(5)$ \cite{b3} symmetry and having a minimum at $\gamma=0$ in the exactly separable $X(5)$ one. Several attempts  have been done in this context with a constant mass parameter \cite{b5a,b5,b6} as well as within the deformation dependent mass formalism \cite{b7,b8,b9,b10}. So, several models have been constructed such as for example E(5)-$\beta^4$ with a pure quartic oscillator potential in $\beta$ shape variable \cite{b11,b12} and ES-X(5)-$\beta^2$ with a harmonic oscillator \cite{b7}. Besides, on the basis of the $X(5)$ symmetry, a prolate $\gamma$-rigid version called $X(3)$ \cite{b13} has been developed which  is parameter independent where the infinite square well potential has been used too. Such a symmetry has motivated the issue of X(3)-$\beta^2$ model \cite{b14} with a harmonic oscillator for the $\beta$ potential. Recently, improved versions of the standard  $X(3)$ and $X(5)$ symmetries being called $X(3)$-ML and $X(5)$-ML have been developed with the introduction of the concept of minimal length \cite{b14a}. Not long ago, an interplay between the $\gamma$-rigid and $\gamma$-stable collective motions in the phase transitions between spherical and axially deformed shapes has been established where the coupling of the two types of the $\beta$ collective motion, described here with an infinite square well like in $X(5)$ and $X(3)$ symmetries as well as with the Davidson potential , is realized via a control parameter indicating the degree of the system's rigidity against $\gamma$ oscillations which in turn are represented by  a simple harmonic oscillator around $\gamma=0$ \cite{b15,b16}.\\
In the present work we treat  the same problematic as in \cite{b15,b16} in a different way using
the position-dependent effective mass  formalism with the Davidson potential  for the $\beta$ collective motion and a potential inspired by a harmonic oscillator for describing the $\gamma$-vibrations. This formlism has been used in many fields of physics such as semiconducters \cite{b16a}, quantum dots  \cite{b16b}, quantum liquids  \cite{b16c,b16d,b16e,b16f} and atomic nuclei  \cite{b8,b9,b10}. Also, it has proved to be efficient to reproduce, with excellent precision, the experimental data for energy spectrum of nuclei  \cite{b8,b9,b10}. Indeed, the obtained numerical results for the energy spectrum  and electric quadrupole $E(2)$ transition rates within the present model called ES-X(3)$\cup$X(5)-D are greatly improved in comparison with those of \cite{b15}. This study is extended to other nuclei in  addition to those previously treated in \cite{b15,b16}.  The eigenenergies and eigenfunctions of the problem were obtained in analytical form by means of the asymptotic iteration method (AIM) \cite{b17,b18}. Such a method has proved to be efficient in the treatment of similar problems \cite{b19,b20,b21,b22,b23}.\\
The content of this paper is arranged as follows. In Sec. \ref{s2}, the position dependent mass formalism is briefly described. In Sec. \ref{s3}, we present a  theoretical framework of the interplay between $\gamma$-stable and $\gamma$-rigid collective motions. Analytical expressions for the energy levels and excited-state wave functions are presented in Sec.  \ref{s4}, while the B(E2) transition probabilities are given in Sec. \ref{s5} . Sec. \ref{s6} is devoted to the discussion and  numerical calculations for energy spectra and B(E2) transition probabilities with their comparisons with experimental data as well as with other theoritical models. Basic concepts of the asymptotic iteration method are given in Appendix A, while in Appendix B, we present the  expressions of the  normalization constants. Finally our conclusions are drawn in Sec. \ref{s7} .
\section{Formalism of position-dependent effective mass  }
\label{s2}
It is well known that  when the mass  operator $m(x)$ is position dependent \cite{b24}, it does not commute with the momentum $\vec{p}=-i\hbar\vec{\nabla}$. Due to this reason there are many ways to generalize the usual form of the kinetic energy $\vec{p}^2/2m_0$ , where $m_0$ is a constant mass, in order to obtain a Hermitian operator. So, in order to avert any specific choices, one can use the general form of the Hamiltonian originally proposed by Von Roos \cite{b25} :
\begin{equation}
	H=-\frac{\hbar^2}{4}\left[m^{\delta'}(x)\nabla m^{\kappa'}(x)\nabla m^{\lambda'}(x)+m^{\lambda'}(x)\nabla m^{\kappa'}(x)\nabla m^{\delta'}(x)\right]+V(x)
	\label{E1}
\end{equation}
where $V(x)$ is the relevant potential and the parameters $\delta',\lambda'$ and $\kappa'$ are constrained by the condition : $\delta'+\kappa'+\lambda'=-1$.
Assuming a position dependent mass of the form  :
\begin{equation}
	m(x)=m_0M(x),\ M(x)=\frac{1}{f(x)^2}
	\label{E2}
\end{equation}
where $m_0$ is a constant mass, $M(x)$ is a dimensionless position-dependent mass, and $f(x)$ is a deforming function, the Hamiltonian (\ref{E1}) becomes,
\begin{equation}
	H=-\frac{\hbar^2}{4m_0}\left[f^{\delta}(x)\nabla f^{\kappa}(x)\nabla f^{\lambda}(x)+f^{\lambda}(x)\nabla f^{\kappa}(x)\nabla f^{\delta}(x)\right]+V(x)
	\label{E3}
\end{equation}
with $\delta+\kappa+\lambda=2$. This Hamiltonian can also be put into the form \cite{b24} :
\begin{equation}
	H=-\frac{\hbar^2}{2m_0}\sqrt{f(x)}\nabla f(x)\nabla \sqrt{f(x)}+V_{eff}(x)
	\label{E4}
\end{equation}
with  
\begin{equation}
	V_{eff}(x)=V(x)+\frac{\hbar^2}{2m_0}\left[\frac{1}{2}\left(1-\delta-\lambda\right)f(x)\nabla^2f(x)+\left(\frac{1}{2}-\delta\right)\left(\frac{1}{2}-\lambda\right)\left[\nabla f(x)\right]^2\right]
	\label{E5}
\end{equation}
\section{Theoretical framework}
\label{s3}
The interplay between $\gamma$-stable and $\gamma$-rigid collective motions, the presence of a deformation-dependent mass term, is achieved by considering the Hamiltonian,
\begin{equation}
	H=\chi \hat{T}_r+(1-\chi)\hat{T}_s+V_{eff}(\beta,\gamma)
		\label{E6}
\end{equation}
where the kinetic energy operator  \cite{b8} associated with a prolate $\gamma$-rigid nucleus reads
\begin{eqnarray}
	\hat{T}_s=\frac{\hbar^2}{2B_m}  \Big(-\frac{\sqrt{f}}{\beta^4}\frac{\partial}{\partial\beta} {\beta^4f}\frac{\partial}{\partial\beta}\sqrt{f}- \frac{f^2}{\beta^2\sin3\gamma}\frac{\partial}{\partial\gamma}\sin3\gamma\frac{\partial}{\partial\gamma}+
	\frac{f^2}{4\beta^2}\sum_ {k=1,2,3}\frac{Q_{k}^{2}}{\sin^2(\gamma-\frac{2}{3}\pi k)} \Big)
		\label{E7}
\end{eqnarray}
while in the $\gamma$-stable case, it is expressed as
\begin{eqnarray}
	\hat{T}_r=\frac{\hbar^2}{2B_m}  \Big( \frac{\sqrt{f}}{\beta^2}\frac{\partial}{\partial\beta} {\beta^2f}\frac{\partial}{\partial\beta}\sqrt{f}-\frac{f^2Q^2}{3\beta^2} \Big)
		\label{E8}
\end{eqnarray}
with,
\begin{equation}
	V_{eff}(\beta,\gamma) =U(\beta,\gamma)+\frac{\hbar^2}{2B_m}  \Big( \frac{1}{2}(1-\delta-\lambda)f\bigtriangledown^2f
	+(\frac{1}{2}-\delta)(\frac{1}{2}-\lambda)(\bigtriangledown f)^{2} \Big)
		\label{E9}
\end{equation}
The variables $\beta$ and $\gamma$ are the usual collective coordinates ( $\beta$ being a deformation coordinate measuring departure from spherical shape, and $\gamma$ being an angle measuring departure from axial symmetry), while $Q_k (k = 1, 2, 3)$ are the components of angular momentum in the intrinsic frame and $B_m$ is the mass parameter, which is usually considered as being a constant. The parameter $\chi\in[0,1[$  is a parameter that indicates the degree of the system's rigidity against $\gamma$ vibrations.\\
In order to achieve exact separation of variables, the reduced potential with the following form :
\begin{equation}
	u(\beta,\gamma)=\frac{2B_m}{\hbar^2}U(\beta,\gamma)=v(\beta)+(1-\chi)\frac{f^2}{\beta^2}w(\gamma)
	\label{E10}
\end{equation}
has been considered and adapted for the present problem.
Using the factorized wave function  $\Psi(\beta,\gamma,\Omega)=\xi(\beta)\varphi(\gamma,\Omega)$ ($\Omega=(\theta_1,\theta_2,\theta_3)$) we can separate the collective Schr\"{o}dinger equation corresponding to the Hamiltonian (\ref{E6}) into two parts:\\
\begin{itemize}
	\item (a) The radial part,
	\begin{eqnarray}
\bigg[\frac{\sqrt{f}}{f^2\beta^{4-2\chi}}\frac{\partial}{\partial\beta}f^2\beta^{4-2\chi}\frac{\partial}{\partial\beta}\sqrt{f}&-\frac{\left(\frac{3}{4}+\delta\lambda-\lambda-\delta\right)f'^2}{f^2}-\frac{\left(1-\delta-\lambda\right)f'}{f\beta} \nonumber \\&-\frac{\left(1-\delta-\lambda\right)f''}{2f} +\frac{\left(\varepsilon-v(\beta)\right)}{f^2}-\frac{W}{\beta^2}\bigg]\xi(\beta)=0
		\label{E11}
	\end{eqnarray}
	\item (b) The angular part,
	\begin{eqnarray}
		\bigg[(1-\chi)\left(-\frac{1}{ sin 3\gamma}\frac{\partial}{\partial\gamma}sin3\gamma\frac{\partial}{\partial\gamma}+\frac{1}{4}\cdot\sum_{k=1}^{3}\frac{Q_k^2}{sin^2\left(\gamma-\frac{2}{3}\pi k\right)}\right)+\frac{\chi}{3}Q^2\nonumber\\ \hspace{3cm}+(1-\chi)\cdot w(\gamma)\bigg]\varphi(\gamma,\Omega)=W\cdot\varphi(\gamma,\Omega)
		\label{E12}
	\end{eqnarray}
\end{itemize}
As pointed out in Ref. \cite{b4}, when the potential has a minimum around $\gamma=0$ , one can write the angular momentum term of Eq. (\ref{E12}) in the form
\begin{eqnarray}
	\sum_{k=1}^{3}\frac{Q_k^2}{sin^2\left(\gamma-\frac{2}{3}\pi k\right)}\approx &\frac{4}{3}\cdot\left(Q_1^2+Q_2^2+Q_3^2\right)-Q_3^2\cdot\left(\frac{1}{sin^2\gamma}-\frac{4}{3}\right)\nonumber\\
	&=\frac{4}{3}\cdot Q^2-Q_3^2\cdot\left(\frac{1}{sin^2\gamma}-\frac{4}{3}\right)
	\label{E13}
\end{eqnarray}
In the same context, the $\gamma$ and angular variables can also be separated by assuming the product state $\varphi(\gamma,\Omega)=\eta(\gamma)D_{M K}^L(\Omega) $,
where we have introduced the Wigner functions $D_{M K}^L(\Omega)$  associated with the total angular momentum $L$ and its projections on the body-fixed and laboratory-fixed $z$ axis, $M$ and $K$, respectively. Averaging the approximated equation (\ref{E12}) on this product state, one obtains the following differential equation for the $\gamma$ shape variable,
\begin{equation}
	\left[-\frac{1}{sin3\gamma}\frac{\partial}{\partial\gamma}sin3\gamma\frac{\partial}{\partial\gamma}+\frac{K^2}{4sin^2\gamma}+w(\gamma)\right]\eta(\gamma)=\epsilon_{\gamma}\eta(\gamma)
	\label{E14}
\end{equation}
with
\begin{equation}
	\epsilon_{\gamma}=\frac{1}{(1-\chi)}\cdot\left(W-\frac{L(L+1)-K^2\left(1-\chi\right)}{3}\right)
	\label{E15}
\end{equation}
\section{Energy spectrum and excited state wave functions}
\label{s4}
The radial equation (\ref{E11}) can be solved for several physical potentials, such as the harmonic oscillator, Davidson  and Kratzer, leading to an excitation spectrum in $\beta$. Furthermore, in this paper, we are going to use as prototype the Davidson potential:
\begin{equation}
	v(\beta)=\beta^2+\frac{\beta_0^4}{\beta^2}
	\label{E16}
\end{equation} 	
where the parameter $\beta_0$ indicates the position of the minimum	of the potential. Notice that, the special case with $\beta_0 = 0$ corresponds to the simple harmonic oscillator. The radial equation (\ref{E11}) can be written as :
\begin{eqnarray}
	\frac{d^2\xi(\beta)}{d\beta}+&\left[\frac{2\left(2-\chi\right)}{\beta}+\frac{2f'}{f}\right]\frac{d\xi(\beta)}{d\beta}+\bigg(\frac{\left(\lambda+\delta-\lambda\delta-\frac{1}{2}\right)f'^2}{f^2}+\frac{\left(1+\lambda+\delta\right)f'}{f\beta}\nonumber\\&+\frac{\left(\delta+\lambda\right)f''}{2f}+\frac{\left(\varepsilon-v(\beta)\right)}{f^2}-\frac{W}{\beta^2}\bigg)\xi(\beta)=0
	\label{E17}
\end{eqnarray}
To derive the eigenvalues and eigenfunctions of this equation  for the potential shown in Eq.(\ref{E16}), we use the AIM (see Appendix A). For this purpose, we  consider for the deformation function the special form \cite{b8,b10} :
\begin{equation}
f(\beta)=1+a\beta^2,\ a\ll 1
\label{E18}
\end{equation}
Bearing in mind the  boundary conditions for the radial wave function, we propose the following ansatz:
\begin{equation}
\xi(\beta)=\beta^{\omega_1}\left(1+a\beta^2\right)^{\omega_2} F_{n_{\beta}}(\beta)
\label{E19}
\end{equation}
with,
\begin{equation}
\omega_1=\chi-\frac{3}{2}+\sqrt{\frac{\left(9+4\chi(\chi-3)+4(\beta_0^4+W)\right)}{4}}
\label{E20}
\end{equation}
\begin{equation}
\omega_2=-\frac{1}{2}+\sqrt{\frac{1+a\varepsilon}{4a^2}+\frac{\beta_0^4+4(\lambda\delta-\lambda-\delta)+3}{4}}
\label{E21}
\end{equation}
Substituting this form of the radial wave function, into Eq. (\ref{E17}) yields,
\begin{eqnarray}
\frac{d^2}{dz^2}F_{n_{\beta}}(z)=-&\left[\frac{\left(\omega_1+2\omega_2-\chi-\frac{9}{2}\right)z+\omega_1-\chi+\frac{5}{2}}{z(1+z)}\right]\frac{d}{dz}F_{n_{\beta}}(z)\nonumber \\&-\left[\frac{\left(\frac{1}{4}(\omega_1+2\omega_2+7)(\omega_1+2\omega_2)-\frac{1}{2}\chi(\omega_1+2\omega_2+1)+Q\right)}{z(1+z)}\right]F_{n_{\beta}}(z)
\label{E22}
\end{eqnarray}
where we have introduced a new variable $z=a\beta^2$, with
\begin{equation}
Q=\frac{7}{4}(\lambda+\delta)-\lambda\delta-\frac{1+Wa^2}{4a^2}
\label{E23}
\end{equation}
The first and the second terms  in square brackets on the right-hand side of Eq. (\ref{E22}) represent $\lambda_0$ and $s_0$ of Eq. (\ref{EE1}), respectively. After calculating $\lambda_n$ and $s_n$, by means of the recurrence relations of Eq. (\ref{EE4}), we get the generalized formula between $\omega_1$ and $\omega_2$ from the roots of the condition (\ref{EE5}) (see Appendix A) :
\begin{equation}
\omega_2=\frac{1}{2}\chi-\frac{1}{2}\omega_1-\frac{(7+4n_{\beta})}{4}-\frac{1}{4}\sqrt{4\chi\left(\chi-5\right)+16Q+49},\ n_{\beta}=0,1,2,\cdots .
\label{E24}
\end{equation}
Once the expressions of $\omega_1$ and $\omega_2$ are obtained, they are
substituted into Eq. (\ref{E24}). So, we get :
\begin{eqnarray}
\varepsilon_{L,n_{\beta},n_{\gamma},K}=& \left(2n_{\beta}+1+\Lambda\right)\left(\sqrt{4a^2\chi(\chi-5)+P}\right)
\nonumber\\
&+ a\bigg(4n_{\beta}\left(n_{\beta}+1+\Lambda\right)+2\chi(\chi-4)-3\left(\lambda+\delta\right)+2(\Lambda+W)+\frac{25}{2}\bigg)
\label{E25}
\end{eqnarray}
with
\begin{equation}
\Lambda=\sqrt{\frac{9+4\chi(\chi-3)}{4}+\beta_0^4+W}, \ P=\left(16\lambda\delta-28(\lambda+\delta)+49\right)a^2+4(1+a^2W)
\label{E26}
\end{equation}
The eigenfunctions corresponding to eigenvalues (\ref{E24}) are obtained in terms of  hypergeometrical functions,
\begin{equation}
F_{n_{\beta}}(z)=C_1\cdot{}_2F_1\left(-n_{\beta},-n_{\beta}-\mu;-\mu-\eta-2n_{\beta};1+z\right)
\label{E27}
\end{equation}
where ${}_2F_1$ represents the hypergeometrical functions with the parameters $\mu$ and $\eta$ given by,
\begin{equation}
\mu=-n_{\beta}-\omega_2-\frac{1}{2}\omega_1+\frac{1}{2}\chi-\frac{7}{2},\
\eta=\omega_1-\chi+\frac{3}{2}
\label{E28}
\end{equation}
Finally, the excited-state wave functions  are obtained,
\begin{equation}
\xi(\beta)=N_{n_{\beta}}\cdot\beta^{\omega_1}\left(1+a\beta^2\right)^{\omega_2} {}_2F_1\left(-n_{\beta},-n_{\beta}-\mu;-\mu-\eta-2n_{\beta};1+a\beta^2\right)
\label{E29}
\end{equation}
where $N_{n_{\beta}}$ is a normalization constant for the radial wave function. This normalization constant is calculated  in Appendix B.\\
As to the angular equation (\ref{E14}) for the $\gamma$ variable, we use a harmonic oscillator,
\begin{equation}
w(\gamma)= (3c)^2\frac{\gamma^2}{2}
\label{E30}
\end{equation}
where the parameter $c$ defining the string constant of the oscillator is usually referred to as the
stiffness of the $\gamma$ oscillations which have a minimum at $\gamma=0$ for axial nuclei. \\ So, by inserting this potential in Eq. (\ref{E14}) and  applying a harmonic approximation for the trigonometric functions centered in $\gamma=0$, one gets the following differential equation :
\begin{equation}
\left[-\frac{1}{\gamma}\frac{\partial}{\partial\gamma}\gamma\frac{\partial}{\partial\gamma}+\frac{K^2}{4\gamma^2}+(3c)^2\frac{\gamma^2}{2}\right]\eta(\gamma)=\varepsilon_{\gamma}\eta(\gamma)
\label{E31}
\end{equation}
The solutions are readily obtained in terms of the Laguerre polynomials \cite{b4} :
\begin{equation}
\eta_{n_{\gamma},|K|}(\gamma)=N_{n,|K|}\gamma^{|K/2|}\exp{\left(-3c\frac{\gamma^{2}}{2}\right)}L_{n}^{|K/2|}(3c\gamma^{2}),
\label{E32}
\end{equation}
where $N_{n,|K|}$ is a normalization constant and $n=(n_{\gamma}-|K/2|)/2$. The corresponding eigenvalues are
\begin{equation}
\varepsilon_{\gamma}=3c(n_{\gamma}+1),\,\,n_{\gamma}=0,1,2,...,
\label{E33}
\end{equation}
with $K=0,\pm 2n_{\gamma}$ for $n_{\gamma}$ even and $K=\pm 2n_{\gamma}$ for $n_{\gamma}$ odd, respectively.
Besides, after inserting the expression of the parameter $W$, which is deduced from the equation (\ref{E15}), into equation (\ref{E25}), we finally obtain the generalized formula for  energy sprectrum,
	\begin{eqnarray}
	\varepsilon_{L,n_{\beta},n_{\gamma},K}=& \left(2n_{\beta}+1+\sqrt{\frac{9+4\chi(\chi-3)}{4}+\beta_0^4+W}\right)\left(\sqrt{4a^2\chi(\chi-5)+P}\right)\nonumber
	\\
	&+ a\Bigg[4n_{\beta}\left(n_{\beta}+1+\sqrt{\frac{9+4\chi(\chi-3)}{4}+\beta_0^4+W}\right)+2\chi(\chi-4)\nonumber\\&-3\left(\lambda+\delta\right)+2\left(\sqrt{\frac{9+4\chi(\chi-3)}{4}+\beta_0^4+W}+W\right)+\frac{25}{2}\Bigg]
	\label{E34}
	\end{eqnarray}
with,	
	\begin{eqnarray}
	& W=\frac{L(L+1)-(1-\chi)K^2}{3}+3c(1-\chi)(n_{\gamma}+1)
	\label{E35}
	\\
    & P=\left(16\lambda\delta-28(\lambda+\delta)+49\right)a^2+4(1+a^2W)
    \label{E36}
	\end{eqnarray}
where $n_{\beta}$ represents the principal quantum number of $\beta$ vibrations. It should
be noticed here, that our expressions for the energy spectrum reproduce exactly the results found in the work \cite{b15}  in the $a\rightarrow 0$ limit. On the other hand, some interesting low-lying bands are classified by the quantum numbers $n_{\beta}$, $n_{\gamma}$  and $K$, such as the ground-state band (g.s.) with $n_{\beta}=n_{\gamma}=K=0$, the $\beta$-band with $n_{\beta}=1$, $n_{\gamma}=K=0$ and finally the $\gamma$-band with  $n_{\beta}=0$, $n_{\gamma}=1$, $K=2$.\\
Furthermore, the full solution of the Hamiltonian (\ref{E6}) after a proper normalization and symmetrization reads as :
\begin{eqnarray}
\Psi_{LMKn_{\beta}n_{\gamma}}(\beta,\gamma,\Omega)=&\xi_{L,K,n_{\beta},n_{\gamma}}(\beta)\eta_{n_{\gamma},|K|}(\gamma)\sqrt{\frac{2L+1}{16\pi^{2}(1+\delta_{K,0})}}\nonumber \\& \times\left[D_{MK}^{L}(\Omega)+(-)^{L}D_{M-K}^{L}(\Omega)\right].
\label{E37}
\end{eqnarray}
With this function one can calculate B(E2) electromagnetic transition  rates.
\section{B(E2) Electromagnetic transitions }
\label{s5}
The general expression for the electric quadrupole transition operator for axially deformed nuclei around $\gamma=0$ is given by \cite{b4}
\begin{equation}
	T_{\mu}^{(E2)}=t\beta\left[D_{\mu,0}^{2}(\Omega)\cos{\gamma}+\frac{1}{\sqrt{2}}\left(D_{\mu,2}^{2}(\Omega)+D_{\mu,-2}^{2}(\Omega)\right)\sin{\gamma}\right],
	\label{E38}
\end{equation}
where $t$ is a scaling factor.
Then, the corresponding $B(E2)$ transition rates from an initial to a final state are given by
\begin{equation}
	B(E2; L K \to L'K')= {5\over 16 \pi} { |\langle L' K' || T_{\mu}^{(E2)}
		|| L K \rangle|^2  \over 2L+1}
	\label{E39}
\end{equation}
where the reduced matrix element can be obtained by employing the
Wigner-Eckrat theorem \cite{b26,b27},
\begin{equation}
	\langle L' M' K' | T^{(E2)}_{\mu} | L M K \rangle = {1\over
		\sqrt{2L'+1}} \langle L 2 L' | M \mu M'\rangle \langle L' K' ||
	T^{(E2)} || LK\rangle .
	\label{E40}
\end{equation}
Note that in the case of  $\Delta K = 0$ the transitions are described by the first term of the $E2$ operator (\ref{E38}) and the second term is for $\Delta K= 2$ transitions.
The final result for the $E2$ transition probability is given in a
factorized form \cite{b28}
\begin{equation}
	B(E2; n L n_\gamma K \to n' L' n'_\gamma K') = {5 \over 16\pi} t^2
	G^2 B_{n,L,n',L'}^2
	C_{n_\gamma,K,n'_\gamma,K'}^2,
	\label{E41}
\end{equation}
where $	G=(\langle L 2 L' | K, K'-K, K'\rangle)$ is the Clebsch-Gordan coefficient dictating the angular momentum selection rules, while 	$B_{n,L,n',L'}$ and $C_{n_\gamma,K,n'_\gamma,K'}$ are integrals over the shape variables $\beta$ and $\gamma$ with integration measures,
\begin{eqnarray}
	B_{n,L,n',L'} = \int \beta \xi_n^L(\beta) \xi_{n'}^{L'}(\beta) \beta^{4-2\chi}
	d\beta,
	\label{E42}
	\\
	C_{n_\gamma,K,n'_\gamma,K'} = \int \sin(\gamma) \eta_{n_\gamma,|K|}(\gamma)\eta_{n'_\gamma,|K'|}(\gamma) |\sin 3\gamma| d\gamma,
	\label{E43}
\end{eqnarray}
\section{ Numerical results and discussion}
\label{s6}
\begin{figure}[H]
	\centering
	\rotatebox{0}{\includegraphics[height=60mm]{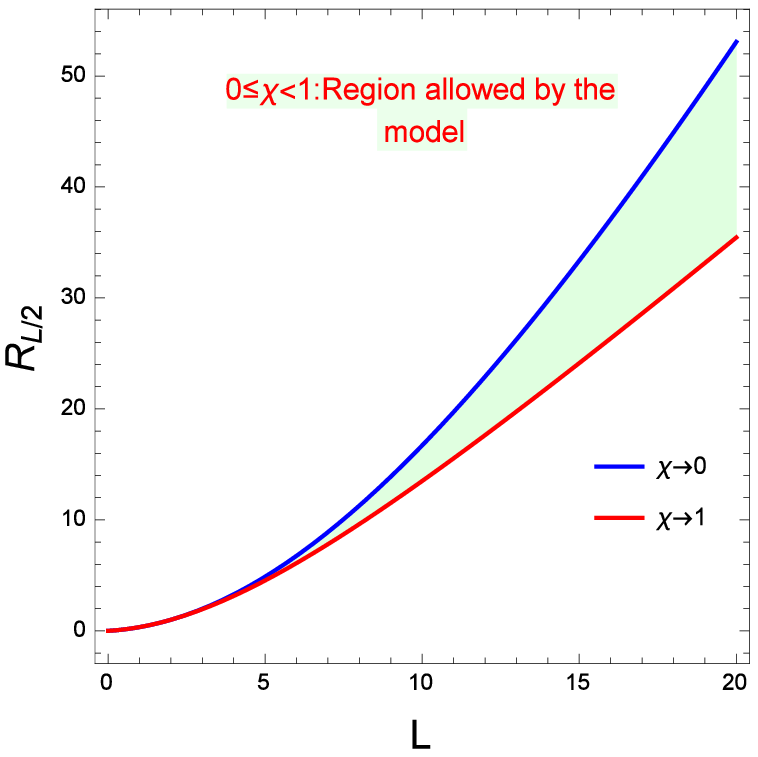}}
	\rotatebox{0}{\includegraphics[height=60mm]{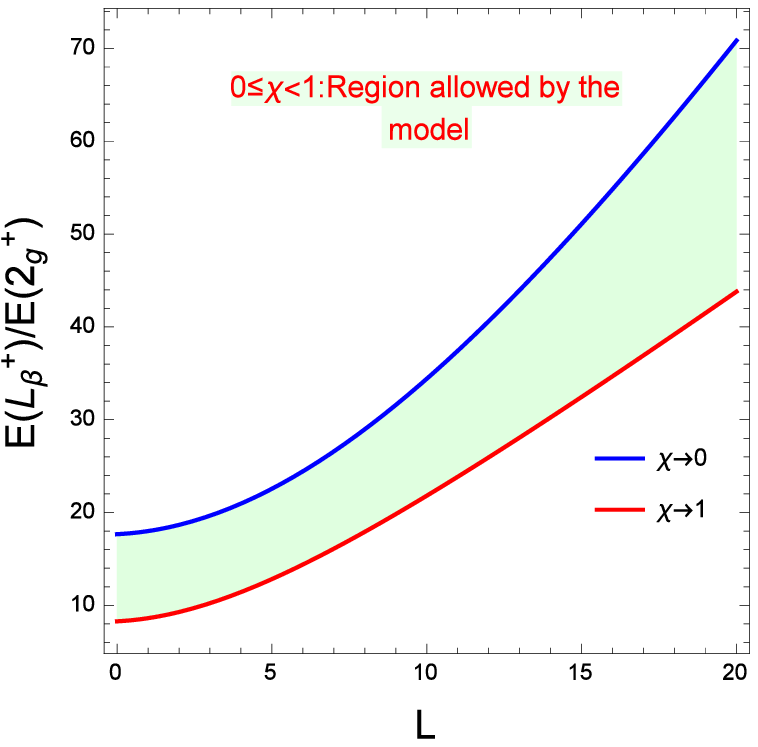}}
	\rotatebox{0}{\includegraphics[height=60mm]{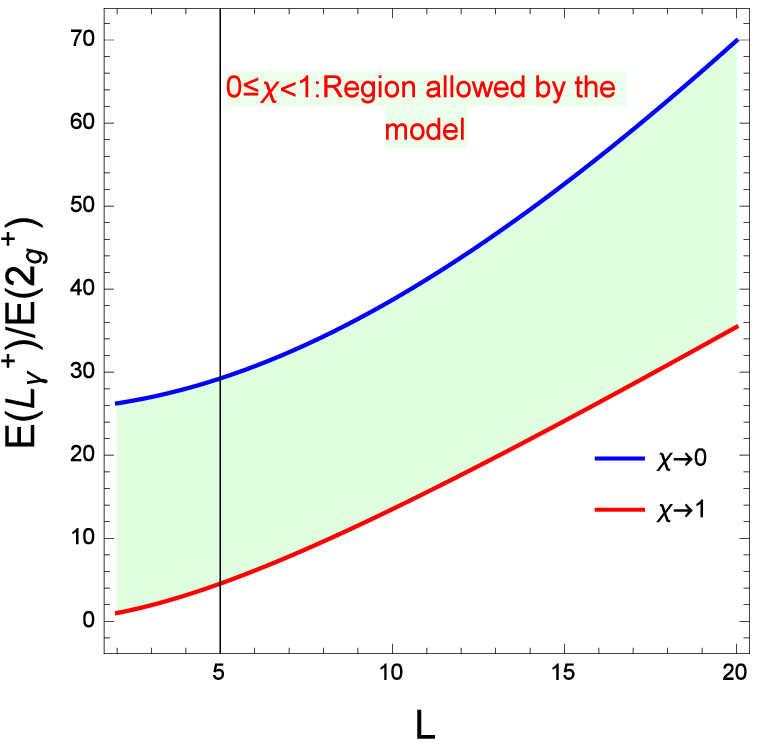}}
	\caption{The region ($0\leq\chi<1$) allowed  by the model associated to $R_{4/2}$, $\beta$ and $\gamma$ band heads normalized to the energy of the first excited state are plotted as function of  angular momentum $L$. }
	\label{Fig1}
\end{figure}

\begin{figure}[H]
	\centering
	\rotatebox{0}{\includegraphics[height=60mm]{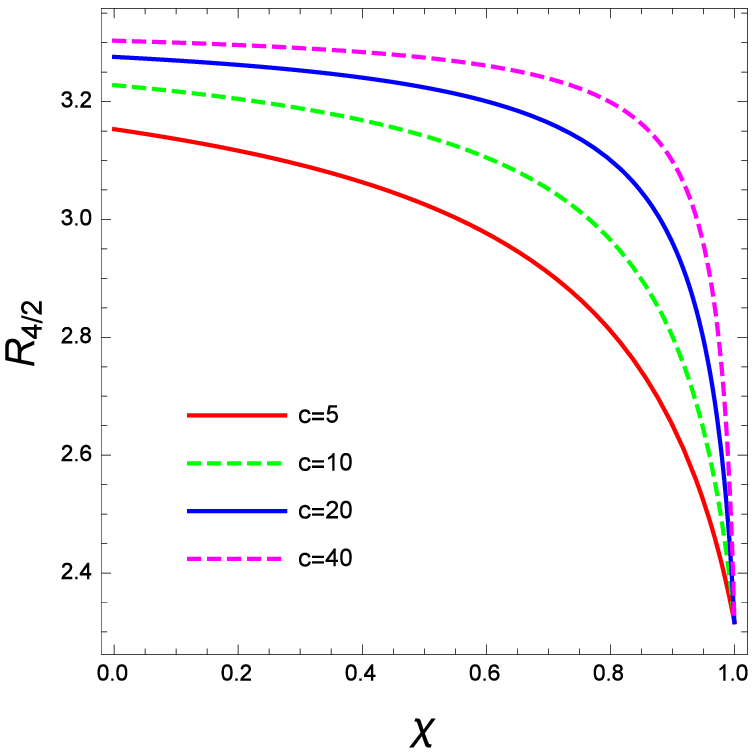}}
	\rotatebox{0}{\includegraphics[height=60mm]{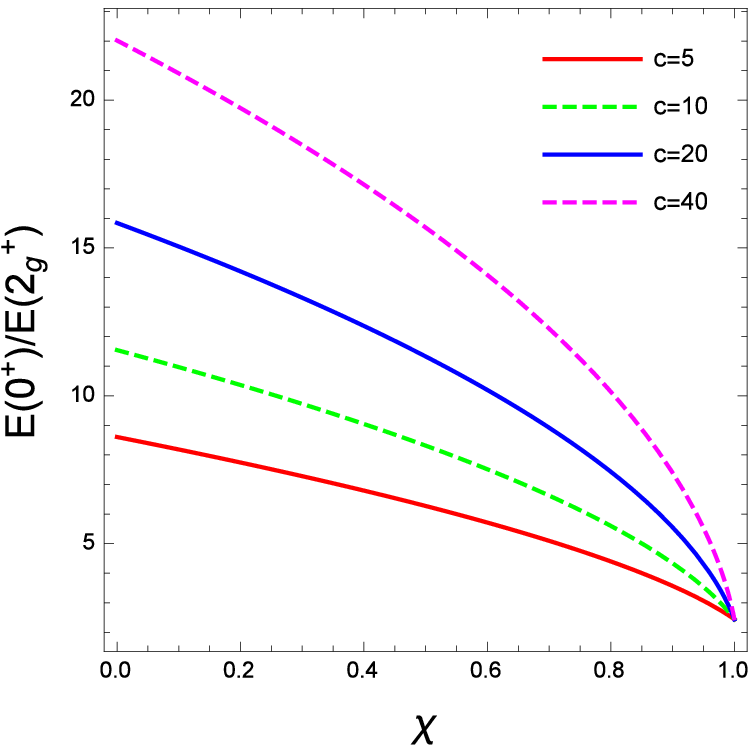}}
	\rotatebox{0}{\includegraphics[height=60mm]{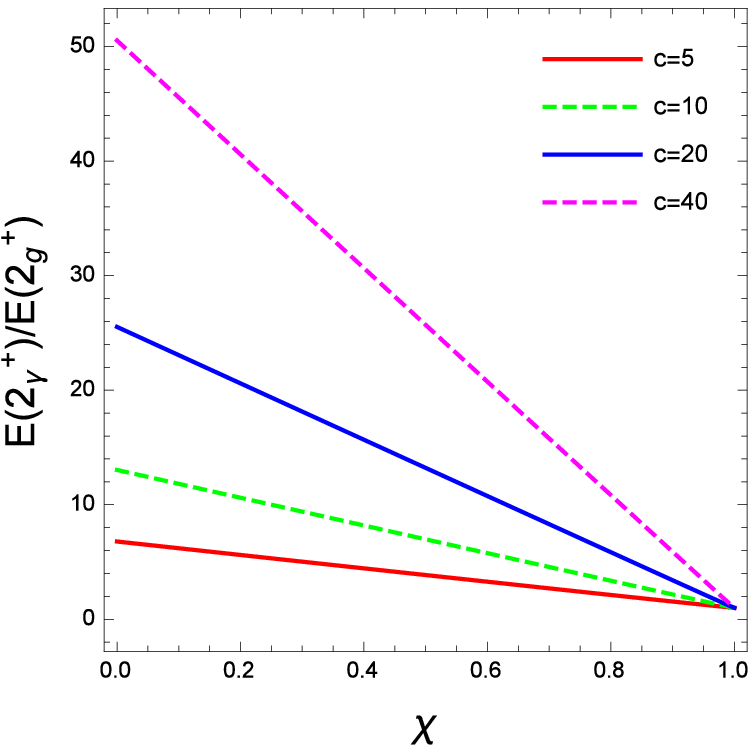}}
	\caption{ $R_{4/2}$, $\beta$ and $\gamma$ band heads normalized to the energy of the first
		excited state are given as functions of the rigidity parameter $\chi$ for $\beta_0=0.8$ and for a few values of the stiffness parameter $c$.}
	\label{Fig2}
\end{figure}

\begin{figure}[h!]
	\centering
	\rotatebox{0}{\includegraphics[height=80mm]{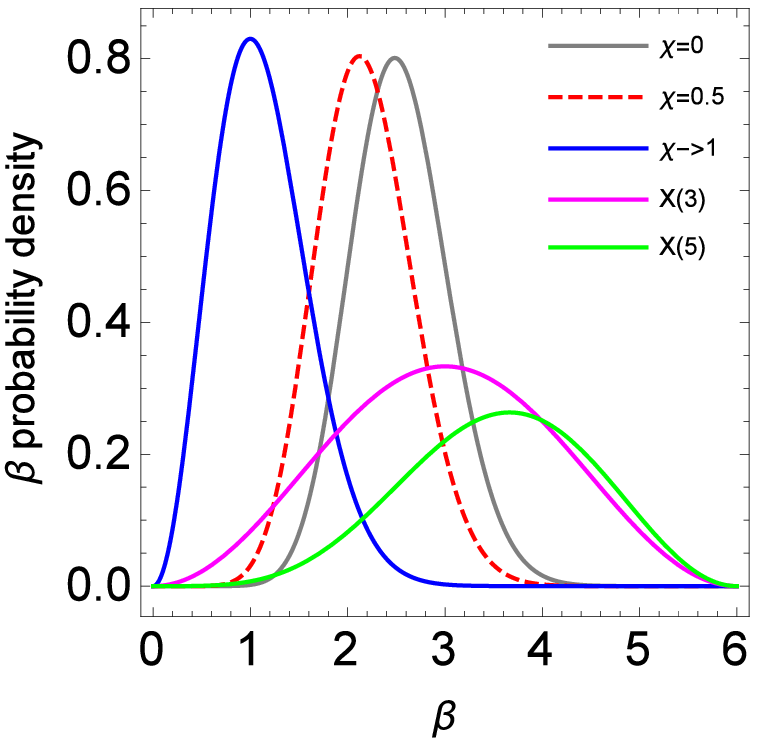}}
	\rotatebox{0}{\includegraphics[height=80mm]{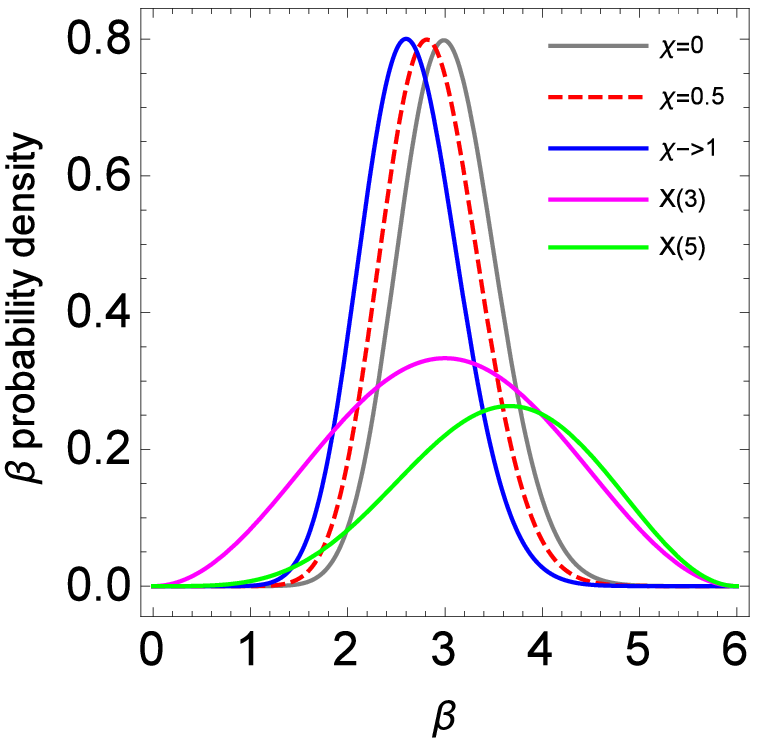}}
	\rotatebox{0}{\includegraphics[height=84mm]{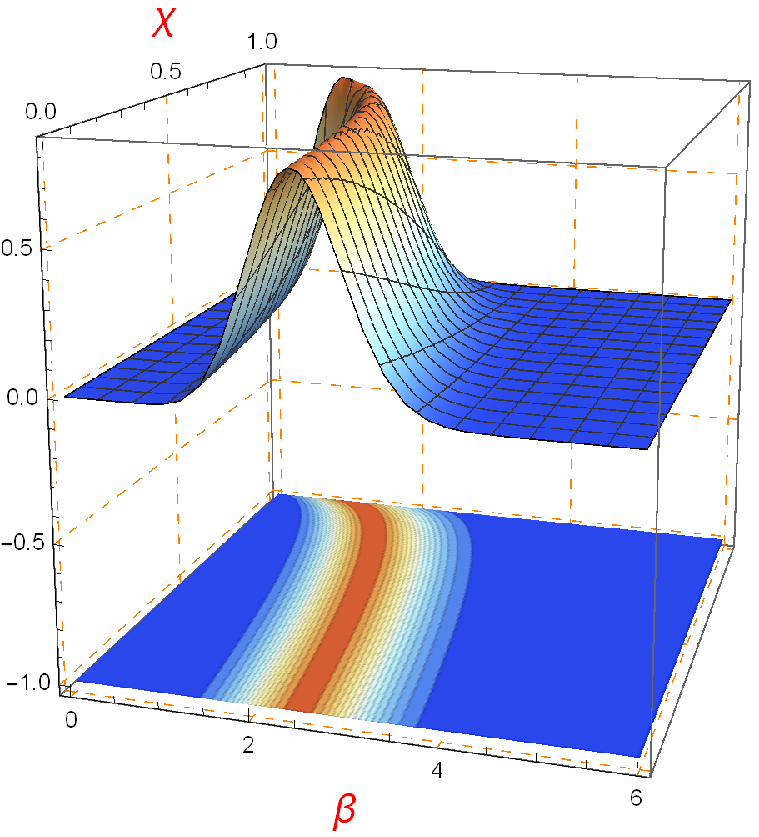}}
	\rotatebox{0}{\includegraphics[height=84mm]{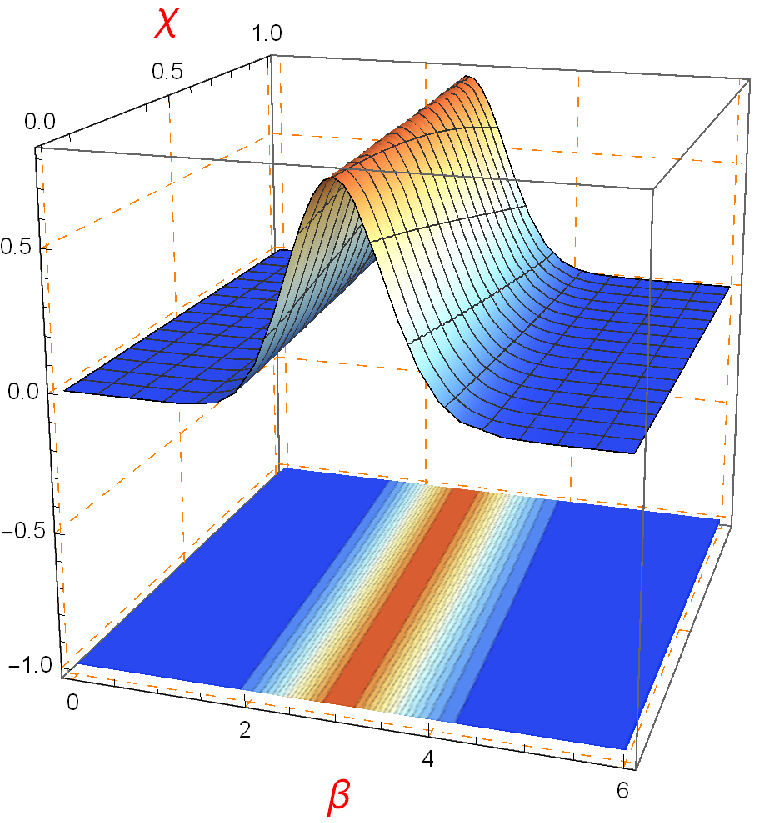}}
	\caption{The ground state $\beta$ probability density with respect to the $d\beta$ integration
		measure in case of the limiting situations $\chi = 0$ and $\chi\rightarrow 1$ for $(c = 10,\ \beta_0=0,\ a=10^{-3})$ and $(c = 10,\ \beta_0=2.5,\ a=10^{-3})$ .
		Probability density $[\xi_{0,0,0,0}(\beta)]^2\beta^{4-2\chi}$ and corresponding lines of constant drawn as functions of $\chi$ and $\beta$ show its
		behaviour in between the previously mentioned limiting cases.}
	\label{Fig3}
\end{figure}

\begin{figure}[h!]
	\centering
	\rotatebox{0}{\includegraphics[height=80mm]{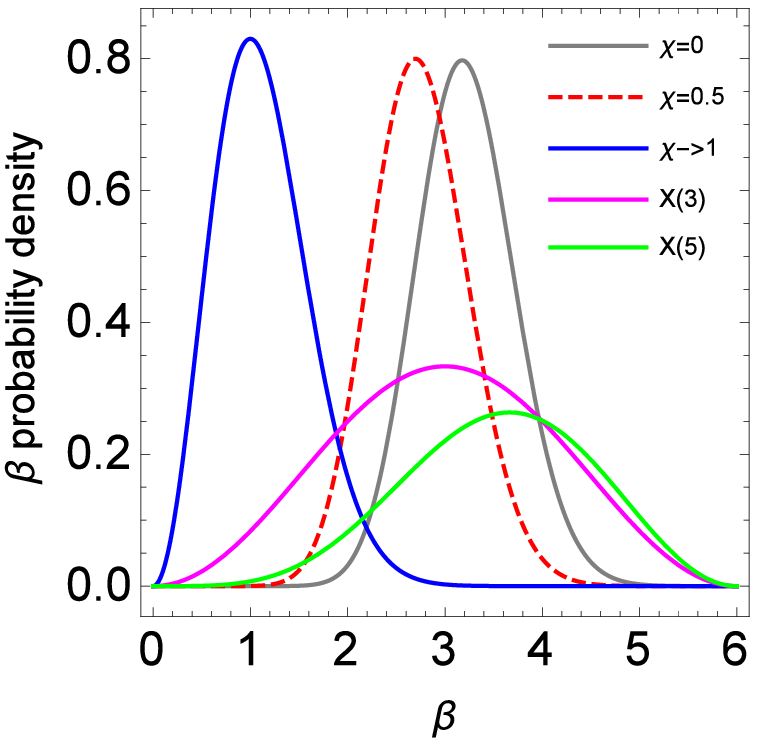}}
	\rotatebox{0}{\includegraphics[height=80mm]{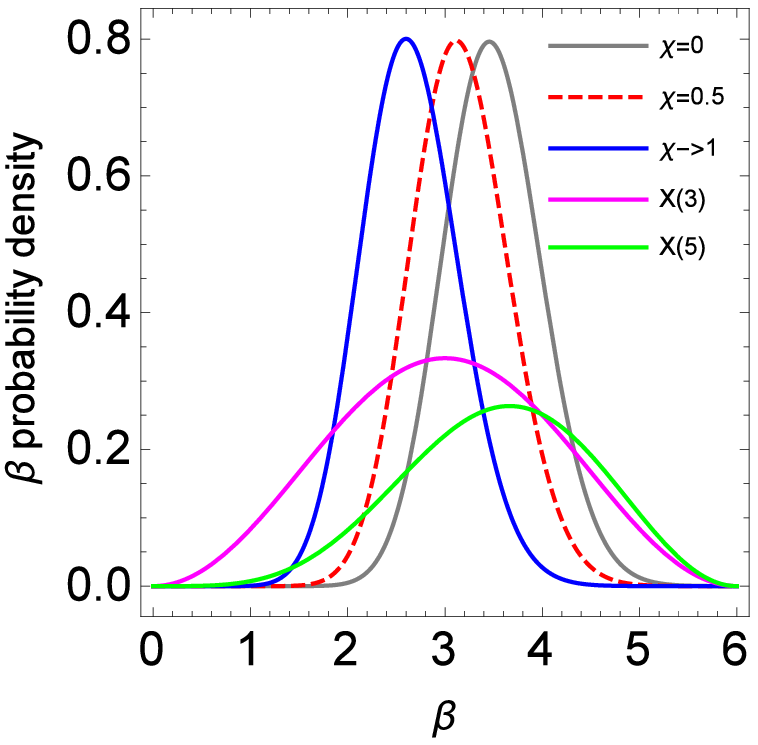}}
	\rotatebox{0}{\includegraphics[height=84mm]{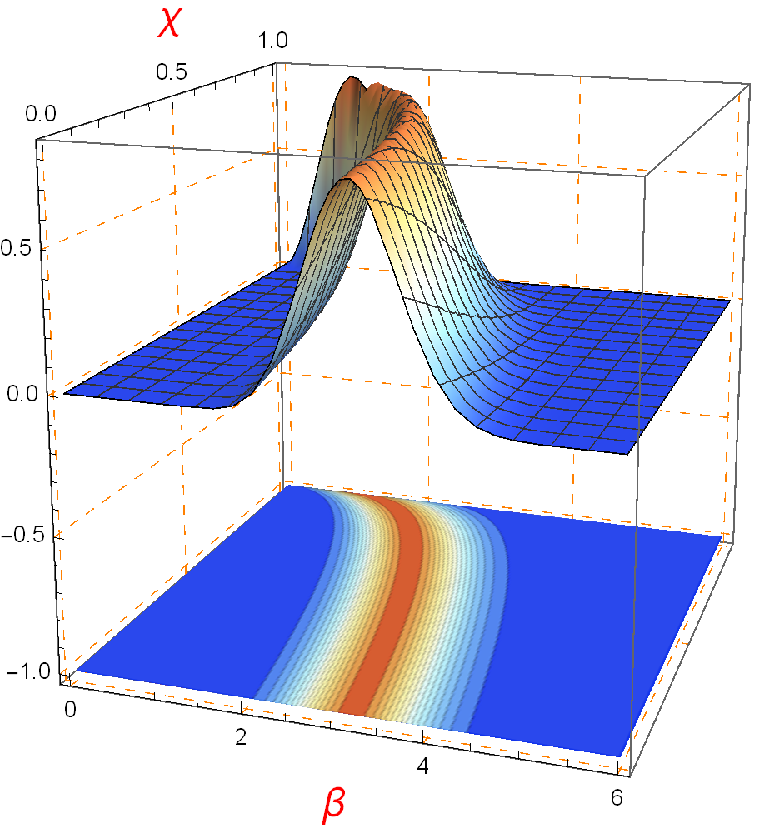}}
	\rotatebox{0}{\includegraphics[height=84mm]{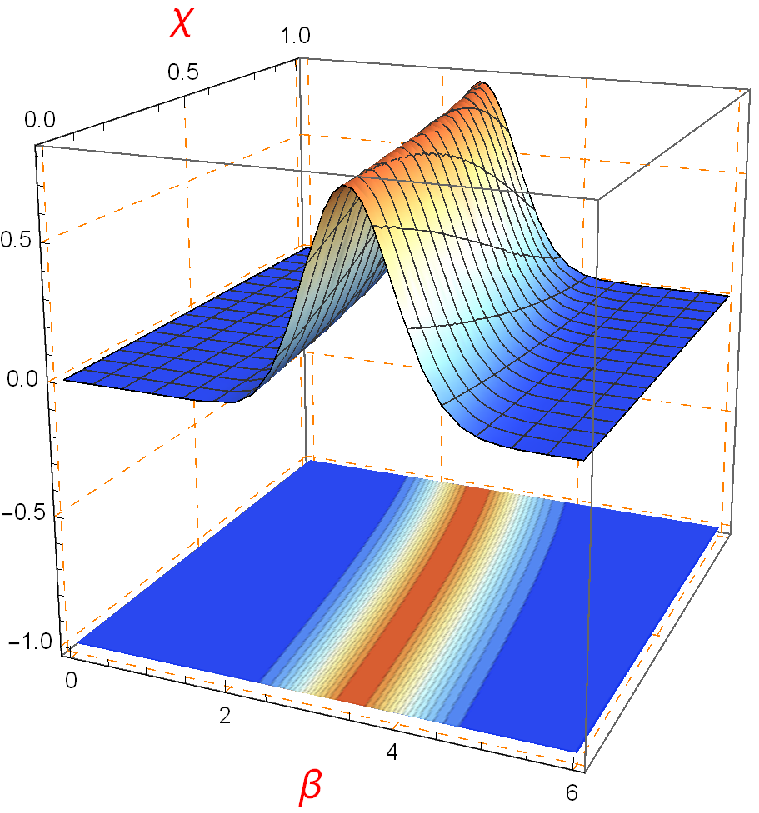}}
	\caption{The same as in Fig. 3, but  for $(c = 30,\ \beta_0=0,\ a=10^{-3})$ and $(c = 30,\ \beta_0=2.5,\ a=10^{-3})$ }
	\label{Fig4}
\end{figure}

\begin{figure}[H]
	\centering
	\rotatebox{0}{\includegraphics[height=100mm]{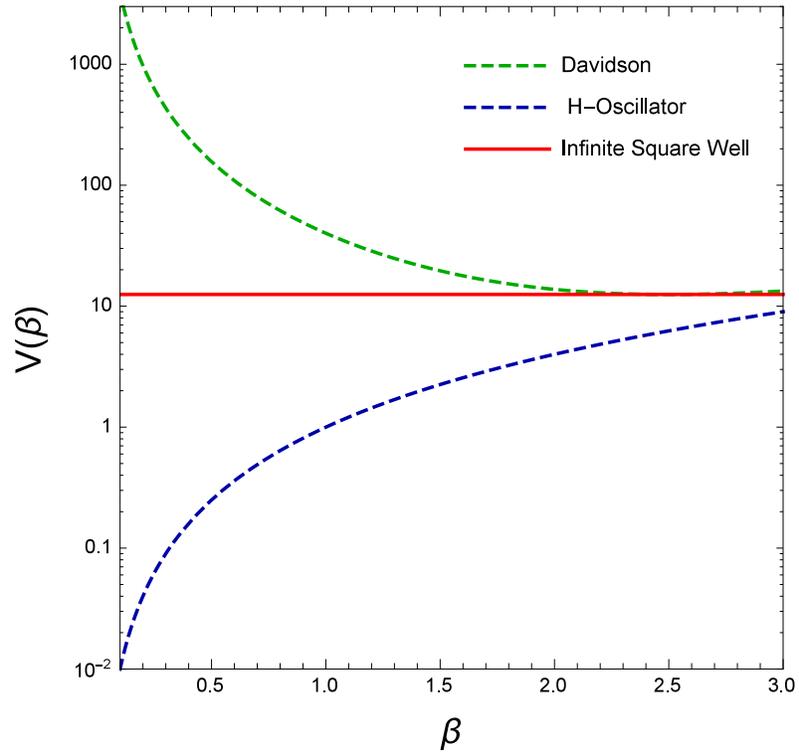}} 	
	\caption{The shape of the Davidson with the minimum $\beta_0=2.5$, Harmonic Osillator and infinite square well potentials are plotted as function of $\beta$ in logarithmic scale.}
	\label{Fig5}
\end{figure}

Before starting any calculations of the energy spectra and transition rates for the axially symmetric prolate deformed nuclei, we recall, as is already mentioned above (Sec.4), a few interesting low-lying bands namely,
\begin{itemize}
	\item (i) The ground state band (g.s) with $n_{\beta}=0$ , $n_{\gamma}=0$, $K = 0$.
	\item (ii) The $\beta$-band with $n_{\beta}=1$ , $n_{\gamma}=0$ , $K = 0$.
	\item (iii) The $\gamma$-band with $n_{\beta}=0$, $n_{\gamma}=1$ , $K = 2$.
\end{itemize}
The region ($0\leq\chi<1$) allowed  by the model associated to $R_{4/2}$, $\beta$ and $\gamma$ band heads normalized to the energy of the first excited state is plotted as a function of  angular momentum $L$ in Fig. \ref{Fig1}. While the variation as a function of $\beta_0$, $c$  and $\chi$ of  the calculated  $R_{4/2}=E(4_g^+)/(2_g^+)$ ratio and the $\beta$ and $\gamma$ band heads normalized to the energy of the first excited state is presented in Fig. \ref{Fig2}.
From Fig. \ref{Fig1}, in the ground state  band one observes a prevalence of $X(3)$ symmetry with a slight contribution from $X(5)$ one at high angular momentum. Indeed, the g.s band is due to rotational behaviour of the nucleus. While in  $\beta$ and $\gamma$ bands, there is a large mixing of both symmetries particularly in the $\gamma$ band. Actually, in both bands the vibrational behaviour is important. From Fig. \ref{Fig2} one can see that the curves corresponding to $R_{4/2}$ and $E(0_\beta^+)/(2_g^+)$ drop down at $\chi\rightarrow1$ to the values of the X(3) model, namely $2.31$ and $2.44$ respectively. These results are coherent with those of \cite{b16}. Also, as in \cite{b16}, at $\chi\rightarrow1$ our model provides the value 1 for the $\gamma$ band head despite the X(3) model excludes the existence of such a  band as it was reported in \cite{b16}. At $\chi=0$, the corresponding values of the ES-X(5) model are also reproduced for all three ratios like in \cite{b16}. Another similarity between the ES-X(3)$\cup$X(5)-D model and the model used in \cite{b16} is the behaviours of the ratios $R_{4/2}$ as well as the excited band heads. Moreover, the height of the observed plateau in the curves of $R_{4/2}$ presents some saturation at higher values of the parameter $c$. So, the maximal value of $R_{4/2}$ = $3.30$ reproduces the well-known rotational limit as in \cite{b16}. Such a behaviour is not observed in the excited $\beta$ and $\gamma$ band heads where the corresponding ratios as well as the harmonic oscillator stiffness parameter $c$ increase.
In figures \ref{Fig3} and \ref{Fig4}, we present the variation of the probability density as a function of the deformation $\beta$ for different values of the Davidson potential minimum $\beta_0$ and  the harmonic oscillator stiffness parameter c . From the upper panels of Fig. \ref{Fig3} one can see that  for $\beta_0=0$  and $c = 10$ (left panel), the probability peaks at both limits $\chi\rightarrow0$ and $\chi\rightarrow1$ are shifted to a lower deformation particularly in the case of $\chi\rightarrow1$. At this minimum value, the Davidson potential is equivalent to the harmonic oscillator one. The latter covers the lower deformation region in respect to the infinite square well as illustrated by Fig. \ref{Fig5}. But, at the limit $\chi\rightarrow0$, the shift is less important than in the case of $\chi\rightarrow1$. This is due to the  $\gamma$-potential which has a dumping effect on the $\beta$-shift. At the minimum $\beta_0=2.5$ (right panel), the Davidson potential tends to the infinite square well form. So, under the effect of the centrifugal part of the Davidson potential, the probability peaks corresponding to both limits $\chi\rightarrow1$ and $\chi\rightarrow0$ as well as the case of $\chi=0.5$, corresponding to the mixing situation of both $\gamma$-rigid and $\gamma$-stable behaviours, are shifted to nearly the middle deformations around $\beta_0 = 2.5$ where the two critical points symmetries X(3) and X(5) are important. But, we have to notice that the probability peak for $\chi\rightarrow0$ has not been shifted more to higher deformations because of the smallness of the harmonic oscillator stiffness parameter $c$ ($c=10$). Hence, the three peaks corresponding to $\chi\rightarrow0 $, $\chi=0.5$ and $\chi\rightarrow1$ are close to each other. In this case, from Fig. \ref{Fig4} (right upper panel) one can see that the probability peak corresponding to the limit $\chi\rightarrow1$ is shifted to the middle region covering the X(3) symmetry. In the equilibrium case between $\gamma$-rigid and $\gamma$-stable behaviours, namely: $\chi=0.5$  the peak is shifted to cover both symmetries X(3) and X(5). While at the limit $\chi\rightarrow0$, the corresponding peak is shifted  more to the X(5) symmetry region. This is due to the  $\gamma$-potential which has increased with the increase of the  parameter c ($c=30$). But, in the case of $\chi=0.5$ there is a balance between both  potential effects. While at the minimum $\beta_0 = 0$ with $c=30$ (left upper panel of Fig. \ref{Fig4}), the peaks corresponding to $\chi=0.5$ and $\chi\rightarrow0 $ are shifted to the middle region under the effect of the $\gamma$- potential but the peak of $\chi\rightarrow1 $ has not moved as in Fig. \ref{Fig3} because of the above cited reasons. Moreover, from the figures \ref{Fig3} and \ref{Fig4} notice  that the amplitude of the different probability peaks obtained in our model are higher than those corresponding to the X(3) and X(5) critical points. This is due to the quadrature measure  $\beta^{4-2\chi}d\beta$. In the lower panels of figures \ref{Fig3}  and \ref{Fig4}, we present the probability density in 3D showing a bell shape with a deformation in the back side where $\chi\rightarrow1$. In this region the symmetry X(3) is prevalent. Such a deformation does not appear in Fig. \ref{Fig3} for $\beta_0=2.5$ and  $c=10$ because, as it was mentioned above, in this situation the probability densities corresponding to the three cases: $\chi\rightarrow 1$, $\chi=0.5$ and $\chi\rightarrow 0$ are close to each other.
Besides, the theoretical  energy spectra and the $B(E2)$ electromagnetic transition rates are predicted with four independent parameters  $\beta_0$, $c$, $a$ and $\chi\in[0,1[$, whose values are obtained from fits to the experimental data by applying a least-squares fitting (rms) procedure. The rms fits of spectra have been performed, using the quality measure
\begin{equation}
	\sigma = \sqrt{ { \sum_{i=1}^N (E_i(exp)-E_i(th))^2 \over
			(N-1)E(2_1^+)^2 } }.
	\label{E44}
\end{equation}
where $N$ denotes the number of states, while $E_i(exp)$ and $E_i(th)$ represent the theoretical and experimental energies of the $i$-th level, respectively. $E(2_1^+)$ is the energy of the first excited level of the g.s. band. Before to start our calculations for the nuclei subject of this study, we have tested our model for some experimental evidences. More precisely, we have calculated the rigidity parameter $\chi$ for $^{172}$Os and  $^{196}$Pt nuclei which have been experimentally proved to be $\gamma$-rigid, i.e. they belong to X(3) symmetry for which $\chi\rightarrow1$.Indeed, we have obtained for both nuclei respectively $\chi=0.9912$ and $\chi=0.987$. Thereafter, our calculations have been first carried out for the even-even isotopes with $A=158-162$ of Gd, the even-even isotopes with $A=160-164$ of Dy and the even-even isotopes with $A=166-170$ of Er. As one can see from Table \ref{Tab1} and Table \ref{Tab2}, our results are in a good agrement with the experimental data and better than those of \cite{b15} for the isotopes of Gd and Dy and than the results of \cite{b16} for $^{166}$Er . Regarding the mixing between $\gamma$-rigid and $\gamma$-stable rotation-vibration which is evaluated by the value of the rigidity parameter $\chi$, we can see that the nucleus $^{168}$Er is the unique good candidate for such a mixing. The corresponding parameter $\chi$ is in the vicinity of 0.5, while all other isotopes have either $\chi\rightarrow0$ or $\chi\rightarrow1$.  The g.s., $\beta$ and $\gamma$ bands of all nuclei are very well reproduced, unlike the work \cite{b15} where the $\beta$ band presents some discrepancies with the experiment like in the work \cite{b16}. Even worse, the $\beta$-band results for $^{160}$Gd in \cite{b16} which have been obtained with square well potential are better than those obtained in \cite{b15} with Davidson potential. Likewise for $^{166}$Er despite this nucleus has not been treated in \cite{b15} . Neverthless, we present, in Table \ref{Tab3}, its energy results obtained with a constant mass parameter. While the isotope $^{166}$Er which has a rigidity parameter $\chi=0.6$ presents some prevalence of the $\gamma$-rigid behaviour in concordance with the obtained result in \cite{b16}, the isotope $^{170}$Er for which the rigidity parameter is $\chi=0.1$ belongs to X(5) symmetry. In addition to the above described nuclei, we have extended our study to the nuclei $^{166}$Yb, $^{168}$Yb, $^{228}$Th and $^{230}$Th.Tables \ref{Tab4} and \ref{Tab5}, show that all these isotopes belong to X(5) symmetry. Concerning the precision of our model's calculations for energy spectra, there is in general a good overlap between our results and the experimental data, particularly for $^{168}$Yb and $^{230}$Th isotopes. In Table \ref{Tab6}, we present our results for probabilities of ground-ground, $\beta$-ground and $\gamma$-ground transitions of nuclei compared to the experimental data and the available results from \cite{b15,b16}. From this table, one can see that our results are in a good agreement with the experiment as well as with those of \cite{b15,b16} for  ground-ground and $\gamma$-ground transitions but more precise than \cite{b15,b16} in  the $\beta$-ground transition.  Moreover, in Table \ref{Tab6} and Table \ref{Tab7}, we compare our results with those of the rigid rotor. In this comparison, our results as well as those of \cite{b15,b16}  are overestimated for the ground-ground transitions, while for the $\gamma$-ground transitions they are identical to the rigid rotor ones.\\
\setlength{\tabcolsep}{6.8pt}
\begin{table}[H]
	\caption{Theoretical results for ground, $\beta$ and $\gamma$ bands energies normalized to the energy of the first excited state $2_{g}^{+}$ are compared with the values taken from Ref. \cite{b15} and with the available experimental data   for $^{158}$Gd \cite{b29}, $^{160}$Gd \cite{b30} and $^{162}$Gd \cite{b31}. The adimensional parameters  $a$, $\beta_0$, $c$ and $\chi$ are  given together with the corresponding deviation $\sigma$ defined by equation (\ref{E44}).}
	\label{Tab1}
	\begin{center}
		{\linespread{0.83}
			\footnotesize
			\begin{tabular}{|c|c|c|c|}
				\hline\noalign{\smallskip}
				&\multicolumn{3}{c|}{$^{158}$\underline{Gd}}\\
				\noalign{\smallskip}\hline\noalign{\smallskip}
				$L$&\underline{Exp.}&\underline{Th.}&\underline{Ref.\cite{b15}}\\
				\noalign{\smallskip}\hline\noalign{\smallskip}
				$2_{g}^{+}$ & 1.00  & 1.00     & 1.00  \\
				$4_{g}^{+}$ & 3.29  & 3.29     & 3.27  \\
				$6_{g}^{+}$ & 6.78  & 6.79     & 6.66  \\
				$8_{g}^{+}$ & 11.37 & 11.38    & 11.00 \\
				$10_{g}^{+}$& 16.98 & 16.95    & 16.11 \\
				$12_{g}^{+}$& 23.47 & 23.38    & 21.84 \\
				$14_{g}^{+}$&       & 30.60    & 28.06 \\
				$16_{g}^{+}$&       & 38.52    & 34.67 \\
				$18_{g}^{+}$&       & 47.09    & 41.58 \\
				$20_{g}^{+}$&       & 56.25    & 48.74 \\
				\noalign{\smallskip}\hline\noalign{\smallskip}
				$0_{\beta}^{+}$ & 15.04 & 14.44 & 14.80 \\
				$2_{\beta}^{+}$ & 15.84 & 15.47 & 15.80 \\
				$4_{\beta}^{+}$ & 17.69 & 17.83 & 18.06 \\
				$6_{\beta}^{+}$ & 20.58 & 21.44 & 21.45 \\
				$8_{\beta}^{+}$ &       & 26.17 & 25.79 \\
				$10_{\beta}^{+}$&       & 31.90 & 30.91 \\
				\noalign{\smallskip}\hline\noalign{\smallskip}
				$2_{\gamma}^{+}$ & 14.93 &  15.10  & 15.34 \\
				$3_{\gamma}^{+}$ & 15.92 &  15.97  & 16.12 \\
				$4_{\gamma}^{+}$ & 17.08 &  17.12  & 17.15 \\
				$5_{\gamma}^{+}$ & 18.63 &  18.54  & 18.42 \\
				$6_{\gamma}^{+}$ & 20.42 &  20.23  & 19.91 \\
				$7_{\gamma}^{+}$ &       &  22.18  & 21.61 \\
				$8_{\gamma}^{+}$ &       &  24.37  & 23.52 \\
				$9_{\gamma}^{+}$ &       &  26.79  & 25.60 \\
				$10_{\gamma}^{+}$&       &  29.45  & 27.85 \\
				$11_{\gamma}^{+}$&       &  32.32  & 30.26 \\
				$12_{\gamma}^{+}$&       &  35.39  & 32.81 \\
				$13_{\gamma}^{+}$&       &  38.67  & 35.50 \\
				$14_{\gamma}^{+}$&       &  42.14  & 38.30 \\
				\noalign{\smallskip}\hline\noalign{\smallskip}
				$\beta_0$  &\multicolumn{3}{c|}{2.37}\\
				$c$  &\multicolumn{3}{c|}{10.63}\\
				$a$  &\multicolumn{3}{c|}{1.803$\cdot10^{-2}$}\\
				$\chi$&\multicolumn{3}{c|}{5.01$\cdot10^{-3}$} \\
				\noalign{\smallskip}\hline\noalign{\smallskip}
				$\sigma$&  & 0.309& 0.601\\
				\hline
			\end{tabular}}
			{\linespread{0.83}
				\footnotesize
				\begin{tabular}{|c|c|c|c|}
					\hline\noalign{\smallskip}
					&\multicolumn{3}{c|}{$^{160}$\underline{Gd}}\\
					\noalign{\smallskip}\hline\noalign{\smallskip}
					$L$&\underline{Exp.}&\underline{Th.}&\underline{Ref.\cite{b15}}\\
					\noalign{\smallskip}\hline\noalign{\smallskip}
					$2_{g}^{+}$ &  1.00  & 1.00    & 1.00  \\
					$4_{g}^{+}$ &  3.30  & 3.30    & 3.29  \\
					$6_{g}^{+}$ &  6.84  & 6.85    & 6.79  \\
					$8_{g}^{+}$ &  11.53 & 11.55   & 11.37 \\
					$10_{g}^{+}$&  17.28 & 17.30   & 16.90 \\
					$12_{g}^{+}$&  24.00 & 24.02   & 23.24 \\
					$14_{g}^{+}$&  31.59 & 31.59   & 30.26 \\
					$16_{g}^{+}$&  39.97 & 39.94   & 37.87 \\
					$18_{g}^{+}$&        & 49.00   & 45.95 \\
					$20_{g}^{+}$&        & 58.71   & 54.44 \\
					\noalign{\smallskip}\hline\noalign{\smallskip}
					$0_{\beta}^{+}$ & 18.33 & 18.19 & 19.34 \\
					$2_{\beta}^{+}$ & 19.08 & 19.22 & 20.34 \\
					$4_{\beta}^{+}$ &       & 21.57 & 22.63 \\
					$6_{\beta}^{+}$ &       & 25.19 & 26.13 \\
					$8_{\beta}^{+}$ &       & 29.99 & 30.71 \\
					$10_{\beta}^{+}$&       & 35.86 & 36.24 \\
					\noalign{\smallskip}\hline\noalign{\smallskip}
					$2_{\gamma}^{+}$ & 13.13 &   13.11   & 13.47\\
					$3_{\gamma}^{+}$ & 14.05 &   14.03   & 14.34 \\
					$4_{\gamma}^{+}$ & 15.25 &   15.24   & 15.50 \\
					$5_{\gamma}^{+}$ & 16.76 &   16.74   & 16.92 \\
					$6_{\gamma}^{+}$ & 18.51 &   18.52   & 18.61 \\
					$7_{\gamma}^{+}$ & 20.58 &   20.57   & 20.55 \\
					$8_{\gamma}^{+}$ & 22.81 &   22.89   & 22.72 \\
					$9_{\gamma}^{+}$ &       &   25.47   & 25.12 \\
					$10_{\gamma}^{+}$&       &   28.29   & 27.73 \\
					$11_{\gamma}^{+}$&       &   31.34   & 30.54\\
					$12_{\gamma}^{+}$&       &   34.62   & 33.53 \\
					$13_{\gamma}^{+}$&       &   38.11   & 36.69 \\
					$14_{\gamma}^{+}$&       &   41.81   & 40.02 \\
					\noalign{\smallskip}\hline\noalign{\smallskip}
					$\beta_0$  &\multicolumn{3}{c|}{2.94}\\
					$c$  &\multicolumn{3}{c|}{40.77}\\
					$a$  &\multicolumn{3}{c|}{1.20$\cdot10^{-2}$}\\
					$\chi$&\multicolumn{3}{c|}{0.790} \\
					\noalign{\smallskip}\hline\noalign{\smallskip}
					$\sigma$&  & 0.0546& 0.768\\
					\hline
				\end{tabular}}
				{\linespread{0.83}
					\footnotesize
					\begin{tabular}{|c|c|c|c|}
						\hline\noalign{\smallskip}
						&\multicolumn{3}{c|}{$^{162}$\underline{Gd}}\\
						\noalign{\smallskip}\hline\noalign{\smallskip}
						$L$&\underline{Exp.}&\underline{Th.}&\underline{Ref.\cite{b15}}\\
						\noalign{\smallskip}\hline\noalign{\smallskip}
						$2_{g}^{+}$ &  1.00  & 1.00    & 1.00  \\
						$4_{g}^{+}$ &  3.30  & 3.30    & 3.30  \\
						$6_{g}^{+}$ &  6.84  & 6.85    & 6.81  \\
						$8_{g}^{+}$ &  11.54 & 11.55   & 11.43 \\
						$10_{g}^{+}$&  17.29 & 17.31   & 17.03 \\
						$12_{g}^{+}$&  24.00 & 24.02   & 23.48 \\
						$14_{g}^{+}$&  31.57 & 31.57   & 30.66 \\
						$16_{g}^{+}$&  39.90 & 39.89   & 38.45 \\
						$18_{g}^{+}$&        & 48.88   & 46.77 \\
						$20_{g}^{+}$&        & 58.48   & 55.53 \\
						\noalign{\smallskip}\hline\noalign{\smallskip}
						$0_{\beta}^{+}$ & 19.93 & 19.88   & 20.46 \\
						$2_{\beta}^{+}$ & 20.84 & 20.89   & 21.46 \\
						$4_{\beta}^{+}$ &       & 23.23   & 23.76 \\
						$6_{\beta}^{+}$ &       & 26.83   & 27.27 \\
						$8_{\beta}^{+}$ &       & 31.59   & 31.89 \\
						$10_{\beta}^{+}$&       & 37.43   & 37.49 \\
						\noalign{\smallskip}\hline\noalign{\smallskip}
						$2_{\gamma}^{+}$ & 12.07 & 12.06  & 12.08 \\
						$3_{\gamma}^{+}$ & 12.99 & 12.98  & 12.98 \\
						$4_{\gamma}^{+}$ & 14.18 & 14.20  & 14.17 \\
						$5_{\gamma}^{+}$ &       & 15.71  & 15.63 \\
						$6_{\gamma}^{+}$ &       & 17.50  & 17.37 \\
						$7_{\gamma}^{+}$ &       & 19.57  & 19.36 \\
						$8_{\gamma}^{+}$ &       & 21.90  & 21.60 \\
						$9_{\gamma}^{+}$ &       & 24.49  & 24.07 \\
						$10_{\gamma}^{+}$&       & 27.31  & 26.76 \\
						$11_{\gamma}^{+}$&       & 30.37  & 29.66 \\
						$12_{\gamma}^{+}$&       & 33.65  & 32.76 \\
						$13_{\gamma}^{+}$&       & 37.15  & 36.03 \\
						$14_{\gamma}^{+}$&       & 40.84  & 39.47 \\
						\noalign{\smallskip}\hline\noalign{\smallskip}
						$\beta_0$  &\multicolumn{3}{c|}{3.05}\\
						$c$  &\multicolumn{3}{c|}{8.86}\\
						$a$  &\multicolumn{3}{c|}{7.60$\cdot10^{-3}$}\\
						$\chi$&\multicolumn{3}{c|}{0.0859} \\
						\noalign{\smallskip}\hline\noalign{\smallskip}
						$\sigma$&  & 0.024& 0.574\\
						\hline
					\end{tabular}}
				\end{center}
				\label{Tab1}
			\end{table}
			\setlength{\tabcolsep}{4.3pt}
			\setlength{\tabcolsep}{6.8pt}
			\begin{table}[H]
				\caption{Theoretical results for ground, $\beta$ and $\gamma$ bands energies normalized to the energy of the first excited state $2_{g}^{+}$ are compared with the values taken from Ref. \cite{b15} and with the available experimental data   for $^{160}$Dy \cite{b30}, $^{162}$Dy \cite{b31} and $^{164}$Dy \cite{b32}. The adimensional parameters  $a$, $\beta_0$, $c$ and $\chi$ are  given together with the corresponding deviation $\sigma$ defined by equation (\ref{E44}).}
		    	\label{Tab2}
				\begin{center}
					{\linespread{0.83}
						\footnotesize
						\begin{tabular}{|c|c|c|c|}
							\hline\noalign{\smallskip}
							&\multicolumn{3}{c|}{$^{160}$\underline{Dy}}\\
							\noalign{\smallskip}\hline\noalign{\smallskip}
							$L$&\underline{Exp.}&\underline{Th.}&\underline{Ref.\cite{b15}}\\
							\noalign{\smallskip}\hline\noalign{\smallskip}
							$2_{g}^{+}$&   1.00  & 1.00  & 1.00  \\
							$4_{g}^{+}$&   3.27  & 3.27  & 3.27  \\
							$6_{g}^{+}$&   6.70  & 6.68  & 6.70  \\
							$8_{g}^{+}$&   11.14 & 11.07 & 11.11 \\
							$10_{g}^{+}$&  16.45 & 16.27 & 16.34 \\
							$12_{g}^{+}$&  22.47 & 22.13 & 22.24 \\
							$14_{g}^{+}$&  28.96 & 28.54 & 28.68 \\
							$16_{g}^{+}$&  35.60 & 35.40 & 35.55 \\
							$18_{g}^{+}$&  42.29 & 42.63 & 42.78 \\
							$20_{g}^{+}$&  49.30 & 50.17 & 50.28 \\
							$22_{g}^{+}$&  56.87 & 57.97 & 58.01 \\
							$24_{g}^{+}$&  65.07 & 66.01 & 65.93 \\
							$26_{g}^{+}$&  73.89 & 74.25 & 74.00 \\
							$28_{g}^{+}$&  83.31 & 82.68 & 82.20 \\
							\noalign{\smallskip}\hline\noalign{\smallskip}
							$0_{\beta}^{+}$& 14.75 & 14.58 & 15.85 \\
							$2_{\beta}^{+}$& 15.55 & 15.59 & 16.85 \\
							$4_{\beta}^{+}$& 17.54 & 17.88 & 19.13 \\
							$6_{\beta}^{+}$&       & 21.31 & 22.55 \\
							$8_{\beta}^{+}$&       & 25.73 & 26.96 \\
							\noalign{\smallskip}\hline\noalign{\smallskip}
							$2_{\gamma}^{+}$& 11.13  & 12.13 & 12.10 \\
							$3_{\gamma}^{+}$& 12.09  & 12.97 & 12.94 \\
							$4_{\gamma}^{+}$& 13.32  & 14.06 & 14.05 \\
							$5_{\gamma}^{+}$& 14.85  & 15.41 & 15.41 \\
							$6_{\gamma}^{+}$& 16.58  & 17.00 & 17.01 \\
							$7_{\gamma}^{+}$& 18.63  & 18.81 & 18.83 \\
							$8_{\gamma}^{+}$& 20.74  & 20.84 & 20.87 \\
							$9_{\gamma}^{+}$& 23.30  & 23.06 & 23.11 \\
							$10_{\gamma}^{+}$& 25.60 & 25.46 & 25.52 \\
							$11_{\gamma}^{+}$& 28.64 & 28.03 & 28.11 \\
							$12_{\gamma}^{+}$& 31.20 & 30.76 & 30.85 \\
							$13_{\gamma}^{+}$& 34.44 & 33.63 & 33.72 \\
							$14_{\gamma}^{+}$& 37.11 & 36.63 & 36.73 \\
							$15_{\gamma}^{+}$& 40.43 & 39.75 & 39.85 \\
							$16_{\gamma}^{+}$& 43.42 & 42.98 & 43.08 \\
							$17_{\gamma}^{+}$& 46.60 & 46.32 & 46.40 \\
							$18_{\gamma}^{+}$& 50.13 & 49.75 & 49.82 \\
							$19_{\gamma}^{+}$& 53.22 & 53.27 & 53.31 \\
							$20_{\gamma}^{+}$& 57.33 & 56.87 & 56.88 \\
							$21_{\gamma}^{+}$& 60.39 & 60.55 & 60.51 \\
							$22_{\gamma}^{+}$& 64.55 & 64.30 & 64.21 \\
							$23_{\gamma}^{+}$& 68.18 & 68.12 & 67.96 \\
							$24_{\gamma}^{+}$&       & 72.00 & 71.76 \\
							$25_{\gamma}^{+}$&       & 75.94 & 75.61 \\
							\noalign{\smallskip}\hline\noalign{\smallskip}
							$\beta_0$  &\multicolumn{3}{c|}{2.31}\\
							$c$  &\multicolumn{3}{c|}{13.92}\\
							$a$  &\multicolumn{3}{c|}{3.82$\cdot10^{-3}$}\\
							$\chi$&\multicolumn{3}{c|}{0.393} \\
							\noalign{\smallskip}\hline\noalign{\smallskip}
							$\sigma$&  & 0.502& 0.636\\
							\hline
						\end{tabular}}
						{\linespread{0.83}
							\footnotesize
							\begin{tabular}{|c|c|c|c|}
								\hline\noalign{\smallskip}
								&\multicolumn{3}{c|}{$^{162}$\underline{Dy}}\\
								\noalign{\smallskip}\hline\noalign{\smallskip}
								$L$&\underline{Exp.}&\underline{Th.}&\underline{Ref.\cite{b15}}\\
								\noalign{\smallskip}\hline\noalign{\smallskip}
								$2_{g}^{+}$&  1.00   & 1.00  & 1.00  \\
								$4_{g}^{+}$&  3.29   & 3.30  & 3.30  \\
								$6_{g}^{+}$&  6.80   & 6.82  & 6.82  \\
								$8_{g}^{+}$&  11.42  & 11.47 & 11.47  \\
								$10_{g}^{+}$& 17.05  & 17.13 & 17.11  \\
								$12_{g}^{+}$& 23.57  & 23.67 & 23.62  \\
								$14_{g}^{+}$& 30.89  & 30.98 & 30.89  \\
								$16_{g}^{+}$& 38.91  & 38.95 & 38.81  \\
								$18_{g}^{+}$& 47.49  & 47.50 & 47.27  \\
								$20_{g}^{+}$& 56.75  & 56.54 & 56.20  \\
								$22_{g}^{+}$& 66.35  & 66.02 & 65.53  \\
								$24_{g}^{+}$& 76.28  & 75.87 & 75.19  \\
								$26_{g}^{+}$&        & 86.05 & 85.14  \\
								$28_{g}^{+}$&        & 96.51 & 95.34  \\
								\noalign{\smallskip}\hline\noalign{\smallskip}
								$0_{\beta}^{+}$& 20.66 & 20.37 & 21.21  \\
								$2_{\beta}^{+}$& 21.43 & 21.37 & 22.21  \\
								$4_{\beta}^{+}$& 23.39 & 23.68 & 24.51  \\
								$6_{\beta}^{+}$&       & 27.22 & 28.03  \\
								$8_{\beta}^{+}$&       & 31.89 & 32.68  \\
								\noalign{\smallskip}\hline\noalign{\smallskip}
								$2_{\gamma}^{+}$ & 11.01 & 10.91 & 11.00 \\
								$3_{\gamma}^{+}$ & 11.94 & 11.82 & 11.91 \\
								$4_{\gamma}^{+}$ & 13.15 & 13.03 & 13.11 \\
								$5_{\gamma}^{+}$ & 14.66 & 14.52 & 14.60 \\
								$6_{\gamma}^{+}$ & 16.42 & 16.29 & 16.37 \\
								$7_{\gamma}^{+}$ & 18.48 & 18.33 & 18.40 \\
								$8_{\gamma}^{+}$ & 20.71 & 20.62 & 20.68 \\
								$9_{\gamma}^{+}$ & 23.28 & 23.15 & 23.20 \\
								$10_{\gamma}^{+}$& 25.88 & 25.91 & 25.94 \\
								$11_{\gamma}^{+}$& 28.98 & 28.89 & 28.90 \\
								$12_{\gamma}^{+}$& 32.52 & 32.07 & 32.06 \\
								$13_{\gamma}^{+}$& 35.45 & 35.44 & 35.40 \\
								$14_{\gamma}^{+}$& 39.00 & 38.99 & 38.92 \\
								$15_{\gamma}^{+}$& 42.57 & 42.70 & 42.60 \\
								$16_{\gamma}^{+}$& 46.30 & 46.58 & 46.44 \\
								$17_{\gamma}^{+}$& 50.08 & 50.60 & 50.41 \\
								$18_{\gamma}^{+}$& 53.84 & 54.76 & 54.51 \\
								$19_{\gamma}^{+}$&       & 59.04 & 58.74 \\
								$20_{\gamma}^{+}$&       & 63.45 & 63.08 \\
								$21_{\gamma}^{+}$&       & 67.97 & 67.52 \\
								$22_{\gamma}^{+}$&       & 72.60 & 72.05 \\
								$23_{\gamma}^{+}$&       & 77.32 & 76.68 \\
								$24_{\gamma}^{+}$&       & 82.14 & 81.39 \\
								$25_{\gamma}^{+}$&       & 87.04 & 86.18 \\
								\noalign{\smallskip}\hline\noalign{\smallskip}
								$\beta_0$  &\multicolumn{3}{c|}{3.02}\\
								$c$  &\multicolumn{3}{c|}{7.63}\\
								$a$  &\multicolumn{3}{c|}{2.34$\cdot10^{-3}$}\\
								$\chi$&\multicolumn{3}{c|}{0.0372} \\							
								\noalign{\smallskip}\hline\noalign{\smallskip}
								$\sigma$&  & 0.258& 0.411\\
								\hline
							\end{tabular}}
							{\linespread{0.83}
								\footnotesize
								\begin{tabular}{|c|c|c|c|}
									\hline\noalign{\smallskip}
									&\multicolumn{3}{c|}{$^{164}$\underline{Dy}}\\
									\noalign{\smallskip}\hline\noalign{\smallskip}
									$L$&\underline{Exp.}&\underline{Th.}&\underline{Ref.\cite{b15}}\\
									\noalign{\smallskip}\hline\noalign{\smallskip}
									$2_{g}^{+}$&  1.00  & 1.00  & 1.00  \\
									$4_{g}^{+}$&  3.30  & 3.30  & 3.30  \\
									$6_{g}^{+}$&  6.83  & 6.84  & 6.84  \\
									$8_{g}^{+}$&  11.50 & 11.52 & 11.52 \\
									$10_{g}^{+}$& 17.19 & 17.23 & 17.23 \\
									$12_{g}^{+}$& 23.79 & 23.85 & 23.85 \\
									$14_{g}^{+}$& 31.20 & 31.27 & 31.27 \\
									$16_{g}^{+}$& 39.32 & 39.38 & 39.37 \\
									$18_{g}^{+}$& 48.08 & 48.08 & 48.07 \\
									$20_{g}^{+}$& 57.39 & 57.29 & 57.28 \\
									$22_{g}^{+}$&       & 66.94 & 66.92 \\
									$24_{g}^{+}$&       & 76.97 & 76.94 \\
									$26_{g}^{+}$&       & 87.31 & 87.28 \\
									$28_{g}^{+}$&       & 97.93 & 97.90 \\
									\noalign{\smallskip}\hline\noalign{\smallskip}
									$0_{\beta}^{+}$& 22.56  & 22.53 & 22.51 \\
									$2_{\beta}^{+}$&        & 23.53 & 23.51 \\
									$4_{\beta}^{+}$&        & 25.83 & 25.81 \\
									$6_{\beta}^{+}$&        & 29.37 & 29.35 \\
									$8_{\beta}^{+}$&        & 34.05 & 34.03 \\
									\noalign{\smallskip}\hline\noalign{\smallskip}	
									$2_{\gamma}^{+}$& 10.38  & 10.25 & 10.23 \\
									$3_{\gamma}^{+}$& 11.28  & 11.18 & 11.16 \\
									$4_{\gamma}^{+}$& 12.48  & 12.40 & 12.38 \\
									$5_{\gamma}^{+}$& 13.96  & 13.92 & 13.90 \\
									$6_{\gamma}^{+}$& 15.75  & 15.72 & 15.69 \\
									$7_{\gamma}^{+}$& 17.75  & 17.78 & 17.76 \\
									$8_{\gamma}^{+}$& 20.04  & 20.11 & 20.09 \\
									$9_{\gamma}^{+}$& 22.55  & 22.69 & 22.67 \\
									$10_{\gamma}^{+}$& 25.33 & 25.50 & 25.48 \\
									$11_{\gamma}^{+}$&       & 28.53 & 28.51 \\
									$12_{\gamma}^{+}$&       & 31.77 & 31.75 \\
									$13_{\gamma}^{+}$&       & 35.21 & 35.19 \\
									$14_{\gamma}^{+}$&       & 38.84 & 38.81 \\
									$15_{\gamma}^{+}$&       & 42.63 & 42.61 \\
									$16_{\gamma}^{+}$&       & 46.59 & 46.56 \\
									$17_{\gamma}^{+}$&       & 50.70 & 50.67 \\
									$18_{\gamma}^{+}$&       & 54.95 & 54.92 \\
									$19_{\gamma}^{+}$&       & 59.33 & 59.29 \\
									$20_{\gamma}^{+}$&       & 63.83 & 63.79 \\
									$21_{\gamma}^{+}$&       & 68.44 & 68.41 \\
									$22_{\gamma}^{+}$&       & 73.16 & 73.12 \\
									$23_{\gamma}^{+}$&       & 77.98 & 77.94 \\
									$24_{\gamma}^{+}$&       & 82.89 & 82.84 \\
									$25_{\gamma}^{+}$&       & 87.88 & 87.83 \\
									\noalign{\smallskip}\hline\noalign{\smallskip}
									$\beta_0$  &\multicolumn{3}{c|}{3.21}\\
									$c$  &\multicolumn{3}{c|}{22.80}\\
									$a$  &\multicolumn{3}{c|}{5.86$\cdot10^{-13}$}\\
									$\chi$&\multicolumn{3}{c|}{0.713} \\							
									\noalign{\smallskip}\hline\noalign{\smallskip}
									$\sigma$&  & 0.077& 0.092\\
									\hline
								\end{tabular}}
							\end{center}
							\label{Tab2}
						\end{table}
						\setlength{\tabcolsep}{4.3pt}
						\setlength{\tabcolsep}{3.pt}
						\begin{table}[H]
							\caption{Theoretical results for ground, $\beta$ and $\gamma$ bands energies normalized to the energy of the first excited state $2_{g}^{+}$ are compared  with the available experimental data \cite{b33} for $^{166}$Er, $^{168}$Er and $^{170}$Er . The adimensional parameters  $c$, $\beta_0$ and $\chi$ are also given together with the corresponding deviation $\sigma$ defined by equation (\ref{E44}).}
						    \label{Tab3}
							\begin{center}
								{\linespread{0.83}
									\footnotesize
									\begin{tabular}{|c|c|c|c|}
										\hline\noalign{\smallskip}
										&\multicolumn{3}{c|}{$^{166}$\underline{Er}}\\
										\noalign{\smallskip}\hline\noalign{\smallskip}
										$L$&\underline{Exp.}&\underline{Th.($a\neq 0$)}&\underline{Th.($a=0$)}\\
										\noalign{\smallskip}\hline\noalign{\smallskip}
										$2_{g}^{+}$&   1.00 & 1.00 & 1.00 \\
										$4_{g}^{+}$&   3.29 & 3.28 & 3.28 \\
										$6_{g}^{+}$&   6.77 & 6.75 & 6.74\\
										$8_{g}^{+}$&   11.31 & 11.26 & 11.24 \\
										$10_{g}^{+}$&  16.75 & 16.66 & 16.61\\
										$12_{g}^{+}$&  22.92 & 22.81 & 22.73\\
										$14_{g}^{+}$&  29.65 & 29.59 & 29.45\\
										$16_{g}^{+}$&  36.84 & 36.88 & 36.67\\
										$18_{g}^{+}$&        & 44.60 & 44.30\\
										$20_{g}^{+}$&        & 52.68 & 52.26\\
										$22_{g}^{+}$&        & 61.06 & 60.49\\
										$24_{g}^{+}$&        & 69.69 & 68.96\\
										\noalign{\smallskip}\hline\noalign{\smallskip}
										$0_{\beta}^{+}$& 18.12 & 17.29& 17.38\\
										$2_{\beta}^{+}$& 18.97 & 18.29& 18.38\\
										$4_{\beta}^{+}$& 20.83 & 20.58& 20.66\\
										$6_{\beta}^{+}$& 23.55 & 24.06& 24.12\\
										$8_{\beta}^{+}$& 27.24 & 28.57& 28.61\\
										$10_{\beta}^{+}$&      & 33.99& 33.99\\
										$12_{\beta}^{+}$&      & 40.15& 40.10\\
										\noalign{\smallskip}\hline\noalign{\smallskip}
										$2_{\gamma}^{+}$& 9.75 & 9.92& 9.98 \\
										$3_{\gamma}^{+}$& 10.67 & 10.81& 10.86\\
										$4_{\gamma}^{+}$& 11.87 & 11.98& 12.03\\
										$5_{\gamma}^{+}$& 13.34 & 13.42& 13.46\\
										$6_{\gamma}^{+}$& 15.09 & 15.12& 15.15\\
										$7_{\gamma}^{+}$& 17.08 & 17.06& 17.08\\
										$8_{\gamma}^{+}$& 19.31 & 19.24& 19.24\\
										$9_{\gamma}^{+}$& 21.73 & 21.63& 21.62\\
										$10_{\gamma}^{+}$& 24.37 & 24.22& 24.19\\
										\noalign{\smallskip}\hline\noalign{\smallskip}
										$\beta_0$ & &\multicolumn{1}{c|}{2.72}& 2.71\\
										$c$ & &\multicolumn{1}{c|}{17.50}& 16.06\\
										$a$ & &\multicolumn{1}{c|}{1.13$10^{-3}$}&-\\
										$\chi$&&\multicolumn{1}{c|}{0.629} & 0.591\\
										\noalign{\smallskip}\hline\noalign{\smallskip}
										$\sigma$& &\multicolumn{1}{c|}{0.380}& 0.390\\
										\hline
									\end{tabular}
									{\linespread{0.83}
										\footnotesize
										\begin{tabular}{|c|c|c|c|}
											\hline\noalign{\smallskip}
											&\multicolumn{3}{c|}{$^{168}$\underline{Er}}\\
											\noalign{\smallskip}\hline\noalign{\smallskip}
											$L$&\underline{Exp.}&\underline{Th.($a\neq0$)}&\underline{Th.($a=0$)}\\
											\noalign{\smallskip}\hline\noalign{\smallskip}
											$2_{g}^{+}$&   1.00 & 1.00 & 1.00\\
											$4_{g}^{+}$&   3.30 & 3.30 & 3.27\\
											$6_{g}^{+}$&   6.87 & 6.86 & 6.66\\
											$8_{g}^{+}$&   11.63 & 11.58 & 11.01\\
											$10_{g}^{+}$&  17.50 & 17.38 & 16.15\\
											$12_{g}^{+}$&  24.35 & 24.19 & 21.90\\
											$14_{g}^{+}$&        & 31.95 & 28.16\\
											$16_{g}^{+}$&        & 40.59 & 34.81\\
											$18_{g}^{+}$&        & 50.07 & 41.77\\
											$20_{g}^{+}$&        & 60.37 & 48.98\\
											$22_{g}^{+}$&        & 71.45 & 56.39\\
											$24_{g}^{+}$&        & 83.31 & 63.97\\
											\noalign{\smallskip}\hline\noalign{\smallskip}
											$0_{\beta}^{+}$& 15.25 & 14.40 & 14.95\\
											$2_{\beta}^{+}$& 15.99 & 15.44 & 15.95\\
											$4_{\beta}^{+}$& 17.68 & 17.85 & 18.22\\
											$6_{\beta}^{+}$& 20.26 & 21.55 & 21.62\\
											$8_{\beta}^{+}$&       & 26.48 & 25.97\\
											$10_{\beta}^{+}$&      & 32.53 & 31.10\\
											$12_{\beta}^{+}$&      & 39.62 & 36.86\\
											\noalign{\smallskip}\hline\noalign{\smallskip}
											$2_{\gamma}^{+}$& 10.29 & 10.30 & 10.80\\
											$3_{\gamma}^{+}$& 11.22 & 11.23 & 11.64\\
											$4_{\gamma}^{+}$& 12.46 & 12.47 & 12.75\\
											$5_{\gamma}^{+}$& 14.00 & 14.01 & 14.10\\
											$6_{\gamma}^{+}$& 15.83 & 15.84 & 15.69\\
											$7_{\gamma}^{+}$& 17.95 & 17.95 & 17.50\\
											$8_{\gamma}^{+}$& 20.35 & 20.34 & 19.52\\
											$9_{\gamma}^{+}$&       & 22.99 & 21.72\\
											$10_{\gamma}^{+}$&      & 25.91 & 24.10\\
											\noalign{\smallskip}\hline\noalign{\smallskip}
											$\beta_0$  & &\multicolumn{1}{c|}{2.73}& 2.36\\
											$c$  &&\multicolumn{1}{|c|}{12.93}& 10.060\\
											$a$  &&\multicolumn{1}{|c|}{2.84$\cdot10^{-2}$}&-\\
											$\chi$& &\multicolumn{1}{|c|}{0.486} & 0.254\\
											\noalign{\smallskip}\hline\noalign{\smallskip}
											$\sigma$& &\multicolumn{1}{c|}{0.417}& 0.863\\
											\hline
										\end{tabular}
										{\linespread{0.83}
											\footnotesize
											\begin{tabular}{|c|c|c|c|}
												\hline\noalign{\smallskip}
												&\multicolumn{3}{c|}{$^{170}$\underline{Er}}\\
												\noalign{\smallskip}\hline\noalign{\smallskip}
												$L$&\underline{Exp.}&\underline{Th.($a\neq0$)}&\underline{Th.($a=0$)}\\
												\noalign{\smallskip}\hline\noalign{\smallskip}
												$2_{g}^{+}$&   1.00 & 1.00 & 1.00 \\
												$4_{g}^{+}$&   3.31 & 3.31 & 3.23\\
												$6_{g}^{+}$&   6.87 & 6.87 & 6.48\\
												$8_{g}^{+}$&   11.64 & 11.65 & 10.52\\
												$10_{g}^{+}$&  17.51 & 17.57 & 15.17\\
												$12_{g}^{+}$&        & 24.59 & 20.26\\
												$14_{g}^{+}$&        & 32.68 & 25.67\\
												$16_{g}^{+}$&        & 41.81 & 31.34\\
												$18_{g}^{+}$&        & 51.98 & 37.19\\
												$20_{g}^{+}$&        & 63.17 & 43.18\\
												$22_{g}^{+}$&        & 75.39 & 49.29\\
												$24_{g}^{+}$&        & 88.62 & 55.49\\
												\noalign{\smallskip}\hline\noalign{\smallskip}
												$0_{\beta}^{+}$& 11.33 & 11.00& 11.65\\
												$2_{\beta}^{+}$& 12.21 & 12.06& 12.65\\
												$4_{\beta}^{+}$& 14.04 & 14.54& 14.88\\
												$6_{\beta}^{+}$&       & 18.37& 18.13\\
												$8_{\beta}^{+}$&       & 23.48& 22.18\\
												$10_{\beta}^{+}$&      & 29.80& 26.82\\
												$12_{\beta}^{+}$&      & 37.28& 31.91\\
												\noalign{\smallskip}\hline\noalign{\smallskip}
												$2_{\gamma}^{+}$& 11.88 & 12.04& 12.84\\
												$3_{\gamma}^{+}$& 12.86 & 12.98& 13.57\\
												$4_{\gamma}^{+}$& 14.34 & 14.23& 14.52\\
												$5_{\gamma}^{+}$& 15.73 & 15.78& 15.69\\
												$6_{\gamma}^{+}$& 17.84 & 17.64& 17.05\\
												$7_{\gamma}^{+}$& 19.81 & 19.80& 18.60\\
												$8_{\gamma}^{+}$&       & 22.23& 20.31\\
												$9_{\gamma}^{+}$&       & 24.96& 22.17\\
												$10_{\gamma}^{+}$&      & 27.97& 24.17\\
												\noalign{\smallskip}\hline\noalign{\smallskip}
												$\beta_0$  & &\multicolumn{1}{c|}{2.32}& 1.14\\
												$c$   & &\multicolumn{1}{|c|}{8.93}& 10.97\\
												$a$  & &\multicolumn{1}{|c|}{5.79$\cdot10^{-2}$}&-\\
												$\chi$& &\multicolumn{1}{|c|}{0.104} & 0.108\\
												\noalign{\smallskip}\hline\noalign{\smallskip}
												$\sigma$& &\multicolumn{1}{c|}{0.195}& 0.939\\
												\hline
											\end{tabular}	
										}}}
									\end{center}
								\end{table}
								\setlength{\tabcolsep}{4.3pt}
								\setlength{\tabcolsep}{7.pt}
								\begin{table}[H]
									\caption{Theoretical results for ground, $\beta$ and $\gamma$ bands energies normalized to the energy of the first excited state $2_{g}^{+}$ are compared  with the available experimental data \cite{b33} for $^{166}$Yb and $^{168}$Yb. The adimensional parameters  $c$, $\beta_0$ and $\chi$ are also given together with the corresponding deviation $\sigma$ defined by equation (\ref{E44}).}
									\label{Tab4}
									\begin{center}
										{\linespread{0.83}
											\footnotesize
											\begin{tabular}{|c|c|c|c|}
												\hline\noalign{\smallskip}
												&\multicolumn{3}{c|}{$^{166}$\underline{Yb}}\\
												\noalign{\smallskip}\hline\noalign{\smallskip}
												$L$&\underline{Exp.}&\underline{Th.($a\neq0$)}&\underline{Th.($a=0$)}\\
												\noalign{\smallskip}\hline\noalign{\smallskip}
												$2_{g}^{+}$&   1.00 & 1.00 & 1.00\\
												$4_{g}^{+}$&   3.23 & 3.24 & 3.25\\
												$6_{g}^{+}$&   6.52 & 6.55 & 6.59\\
												$8_{g}^{+}$&   10.73 & 10.76 & 10.83\\
												$10_{g}^{+}$&  15.69 & 15.72 & 15.77\\
												$12_{g}^{+}$&  21.27 & 21.33 & 21.25\\
												$14_{g}^{+}$&   &  27.48  & 27.16\\
												$16_{g}^{+}$&   &  34.16 & 33.40\\
												$18_{g}^{+}$&      & 41.33 & 39.89\\
												\noalign{\smallskip}\hline\noalign{\smallskip}
												$0_{\beta}^{+}$& 10.19 & 9.75 & 13.47\\
												$2_{\beta}^{+}$& 11.18 & 10.80& 14.47\\
												$4_{\beta}^{+}$& 13.11 & 13.13& 16.72\\
												$6_{\beta}^{+}$& 15.70 & 16.58& 20.06\\
												\noalign{\smallskip}\hline\noalign{\smallskip}
												$2_{\gamma}^{+}$& 9.11 & 9.56 & 11.62\\
												$3_{\gamma}^{+}$& 10.15 & 10.38 & 12.41\\
												$4_{\gamma}^{+}$& 11.36 & 11.46 & 13.46\\
												$5_{\gamma}^{+}$& 12.97 & 12.79 & 14.74\\
												$6_{\gamma}^{+}$& 14.48 & 14.35 & 16.24\\
												$7_{\gamma}^{+}$& 16.65 & 16.12 & 17.95\\
												\noalign{\smallskip}\hline\noalign{\smallskip}
												$\beta_0$ & &\multicolumn{1}{c|}{1.55}& 2.02\\
												$c$  & &\multicolumn{1}{c|}{6.95}& 8.44\\
												$a$  & &\multicolumn{1}{c|}{2.48$\cdot10^{-2}$}&-\\
												$\chi$ & &\multicolumn{1}{c|}{0.0229} & 0.0017\\
												\noalign{\smallskip}\hline\noalign{\smallskip}
												$\sigma$ & &\multicolumn{1}{c|}{0.338}& 0.514\\
												\hline
											\end{tabular}
											{\linespread{0.83}
												\footnotesize
												\begin{tabular}{|c|c|c|c|}
													\hline\noalign{\smallskip}
													&\multicolumn{3}{c|}{$^{168}$\underline{Yb}}\\
													\noalign{\smallskip}\hline\noalign{\smallskip}
													$L$&\underline{Exp.}&\underline{Th.($a\neq0$)}&\underline{Th.($a=0$)}\\
													\noalign{\smallskip}\hline\noalign{\smallskip}
													$2_{g}^{+}$&   1.00 & 1.00 & 1.00 \\
													$4_{g}^{+}$&   3.27 & 3.26 & 3.20\\
													$6_{g}^{+}$&   6.67 & 6.66 & 6.37\\
													$8_{g}^{+}$&   11.06 & 11.03& 10.26\\
													$10_{g}^{+}$&  16.25 & 16.20 & 14.67\\
													$12_{g}^{+}$&  22.07 & 22.06 & 19.44\\
													$14_{g}^{+}$&  28.37 & 28.48 & 24.49\\
													$16_{g}^{+}$&   & 35.39 & 29.72\\
													$18_{g}^{+}$&      & 42.73 & 35.11\\
													\noalign{\smallskip}\hline\noalign{\smallskip}
													$0_{\beta}^{+}$& 13.15 & 12.91& 10.46\\
													$2_{\beta}^{+}$& 14.06 & 13.93 & 11.46 \\
													$4_{\beta}^{+}$& 15.86 & 16.24 & 13.66\\
													$6_{\beta}^{+}$&  & 19.71 & 20.72 \\
													\noalign{\smallskip}\hline\noalign{\smallskip}
													$2_{\gamma}^{+}$& 11.21 & 11.47 & 9.89\\
													$3_{\gamma}^{+}$& 12.16 & 12.30 & 10.62\\
													$4_{\gamma}^{+}$& 13.35 & 13.41 & 11.58\\
													$5_{\gamma}^{+}$& 14.84 & 14.76 & 12.74\\
													$6_{\gamma}^{+}$& 16.47 & 16.35 & 14.10\\
													$7_{\gamma}^{+}$& 18.45 & 18.17 & 15.63 \\
													\noalign{\smallskip}\hline\noalign{\smallskip}
													$\beta_0$ & &\multicolumn{1}{c|}{2.10}& 1.16\\
													$c$ & &\multicolumn{1}{c|}{8.151}& 7.47\\
													$a$ & &\multicolumn{1}{c|}{1.08$\cdot10^{-2}$}&-\\
													$\chi$& &\multicolumn{1}{c|}{3.36$\cdot10^{-3}$}& 5.05$\cdot10^{-3}$\\
													\noalign{\smallskip}\hline\noalign{\smallskip}
													$\sigma$& &\multicolumn{1}{c|}{0.168}& 0.748\\
													\hline
												\end{tabular}
											}}	
											
										\end{center}
									\end{table}
									\setlength{\tabcolsep}{4.3pt}
									\setlength{\tabcolsep}{7.pt}
									\begin{table}[H]
										\caption{Theoretical results for ground, $\beta$ and $\gamma$ bands energies normalized to the energy of the first excited state $2_{g}^{+}$ are compared  with the available experimental data\cite{b33} for $^{228}$Th and $^{230}$Th. The adimensional parameters  $c$, $\beta_0$ and $\chi$ are also given together with the corresponding deviation $\sigma$ defined by equation (\ref{E44}).}
										\label{Tab5}
										\begin{center}
											{\linespread{0.83}
												\footnotesize
												\begin{tabular}{|c|c|c|c|}
													\hline\noalign{\smallskip}
													&\multicolumn{3}{c|}{$^{228}$\underline{Th}}\\
													\noalign{\smallskip}\hline\noalign{\smallskip}
													$L$&\underline{Exp.}&\underline{Th.($a\neq0$)}& \underline{Th.($a=0$)}\\
													\noalign{\smallskip}\hline\noalign{\smallskip}
													$2_{g}^{+}$&   1.00 & 1.00 & 1.00 \\
													$4_{g}^{+}$&   3.23 & 3.26 & 3.26\\
													$6_{g}^{+}$&   6.55 & 6.64 & 6.64\\
													$8_{g}^{+}$&   10.78& 10.97 & 10.96\\
													$10_{g}^{+}$&  15.79& 16.05 & 16.03\\
													$12_{g}^{+}$&  21.46& 21.75 & 21.71\\
													$14_{g}^{+}$&  27.69& 27.94 & 27.85\\
													$16_{g}^{+}$&  34.42& 34.51 & 34.37\\
													$18_{g}^{+}$&  41.69& 41.39 & 41.19\\
													\noalign{\smallskip}\hline\noalign{\smallskip}
													$0_{\beta}^{+}$& 14.40 & 14.24& 14.47\\
													$2_{\beta}^{+}$& 15.14 & 15.25& 15.47\\
													$4_{\beta}^{+}$&  & 17.52& 17.73\\
													$6_{\beta}^{+}$&  & 20.91& 21.11\\
													
													\noalign{\smallskip}\hline\noalign{\smallskip}
													
													$2_{\gamma}^{+}$& 16.78 & 17.05& 17.05\\
													$3_{\gamma}^{+}$& 17.70 & 17.81& 17.81 \\
													$4_{\gamma}^{+}$& 18.89 & 18.80& 18.80\\
													$5_{\gamma}^{+}$& 20.33 & 20.03& 20.03\\
													$6_{\gamma}^{+}$&       & 21.48& 21.47\\
													$7_{\gamma}^{+}$&       & 23.14& 23.12\\
													
													\noalign{\smallskip}\hline\noalign{\smallskip}
													$\beta_0$ & &\multicolumn{1}{c|}{1.77}& 1.80\\
													$c$  & &\multicolumn{1}{c|}{12.87}& 12.89\\
													$a$  & &\multicolumn{1}{c|}{1.42$\cdot10^{-3}$}&-\\
													$\chi$& &\multicolumn{1}{c|}{6.62$\cdot10^{-4}$}& 4.97$\cdot10^{-4}$ \\
													\noalign{\smallskip}\hline\noalign{\smallskip}
													$\sigma$& &\multicolumn{1}{c|}{0.203}& 0.231\\
													\hline
												\end{tabular}
												{\linespread{0.83}
													\footnotesize
													\begin{tabular}{|c|c|c|c|}
														\hline\noalign{\smallskip}
														&\multicolumn{3}{c|}{$^{230}$\underline{Th}}\\
														\noalign{\smallskip}\hline\noalign{\smallskip}
														$L$&\underline{Exp.}&\underline{Th.($a\neq0$)}&\underline{Th.($a=0$)}\\
														\noalign{\smallskip}\hline\noalign{\smallskip}
														$2_{g}^{+}$&   1.00 & 1.00 & 1.00\\
														$4_{g}^{+}$&   3.27 & 3.27 & 3.24\\
														$6_{g}^{+}$&   6.70 & 6.71 & 6.52\\
														$8_{g}^{+}$&   11.17 & 10.17 & 10.63 \\
														$10_{g}^{+}$&  16.54 & 16.52 & 15.38\\
														$12_{g}^{+}$&  22.70 & 22.67 & 20.60\\
														$14_{g}^{+}$&   & 29.54 & 26.20\\
														$16_{g}^{+}$&   &  37.06 & 32.04\\
														$18_{g}^{+}$&      & 45.20 & 38.10\\
														\noalign{\smallskip}\hline\noalign{\smallskip}
														$0_{\beta}^{+}$& 11.93 & 11.56 & 12.23\\
														$2_{\beta}^{+}$& 12.74 & 12.61 & 13.23\\
														$4_{\beta}^{+}$& 14.47 & 14.98 & 15.47\\
														$6_{\beta}^{+}$&  & 18.56 & 18.75\\
														\noalign{\smallskip}\hline\noalign{\smallskip}
														$2_{\gamma}^{+}$& 14.67 & 14.68& 14.80\\
														$3_{\gamma}^{+}$& 15.52 & 15.52& 15.52\\
														$4_{\gamma}^{+}$& 16.61 & 16.62& 16.46\\
														$5_{\gamma}^{+}$&  & 17.98& 17.62\\
														$6_{\gamma}^{+}$& & 19.60& 18.97\\
														$7_{\gamma}^{+}$&  & 21.45& 20.51\\
														\noalign{\smallskip}\hline\noalign{\smallskip}
														$\beta_0$ & &\multicolumn{1}{c|}{1.75}& 0.051\\
														$c$ & &\multicolumn{1}{c|}{11.12} & 12.12\\
														$a$ & &\multicolumn{1}{c|}{2.58$\cdot10^{-2}$} &-\\
														$\chi$& &\multicolumn{1}{c|}{0.0558} & 0.0557\\
														\noalign{\smallskip}\hline\noalign{\smallskip}
														$\sigma$ & &\multicolumn{1}{c|}{0.195}& 0.823\\
														\hline
													\end{tabular}
													
												}}	
											\end{center}
										\end{table}
										\setlength{\tabcolsep}{4.3pt}
										\setlength{\tabcolsep}{3.3pt}
										\begin{table}[H]
											\caption{ The comparison of the theoretical predictions of B(E2) transition probabilities  with available experimental data (upper line) and with  theoretical predictions in \cite{b15,b16}, for $^{158-162}$Gd and $^{160-164}$Dy isotopes  . $\Delta K=0$ transition rates are normalized to the $2_{g}^{+}\rightarrow0^{+}_{g}$ transition, while $\Delta K=2$ transitions to the $2_{\gamma}^{+}\rightarrow0^{+}_{g}$ transition, as in \cite{b26,b28}. The rigid rotor estimations are also presented for reference.}
											\label{Tab6}
											\begin{center}
												\begin{tabular}{|c|llllllllll|}
													\noalign{\smallskip}\hline\noalign{\smallskip}
													Nucleus&$\frac{4^{+}_{g}\rightarrow2^{+}_{g}}{2^{+}_{g}\rightarrow0^{+}_{g}}$& $\frac{6^{+}_{g}\rightarrow4^{+}_{g}}{2^{+}_{g}\rightarrow0^{+}_{g}}$& $\frac{8^{+}_{g}\rightarrow6^{+}_{g}}{2^{+}_{g}\rightarrow0^{+}_{g}}$& $\frac{10^{+}_{g}\rightarrow8^{+}_{g}}{2^{+}_{g}\rightarrow0^{+}_{g}}$& $\frac{12^{+}_{g}\rightarrow10^{+}_{g}}{2^{+}_{g}\rightarrow0^{+}_{g}}$&
													$\frac{14^{+}_{g}\rightarrow12^{+}_{g}}{2^{+}_{g}\rightarrow0^{+}_{g}}$&
													$\frac{2^{+}_{\beta}\rightarrow2^{+}_{g}}{2^{+}_{\beta}\rightarrow0^{+}_{g}}$&
													$\frac{2^{+}_{\beta}\rightarrow4^{+}_{g}}{2^{+}_{\beta}\rightarrow0^{+}_{g}}$&
													$\frac{2^{+}_{\gamma}\rightarrow2^{+}_{g}}{2^{+}_{\gamma}\rightarrow0^{+}_{g}}$& $\frac{2^{+}_{\gamma}\rightarrow4^{+}_{g}}{2^{+}_{\gamma}\rightarrow0^{+}_{g}}$\\
													\noalign{\smallskip}\hline\noalign{\smallskip}
													$^{158}$Gd& 1.46(5)&  & 1.67(16) & 1.72(16) & & & 0.25(6) & 4.48(75) & 1.76(26) & 0.079(14) \\
													This work& 1.46 & 1.66& 1.82 & 1.97 & 2.13 & 2.30 & 1.41 & 2.49 & 1.45 & 0.075\\
													Ref. \cite{b15}&   &  &  &   &  &  &  	1.93  & 6.01  & 1.46  & 0.077 \\
													\noalign{\smallskip}\hline\noalign{\smallskip}
													$^{160}$Gd& & & & & & & & & 1.87(12) &  0.189(29)  \\
													This work & 1.45 & 1.63& 1.77 & 1.89 & 2.01 & 2.14 & 1.42 & 2.52& 1.44 & 0.073\\
													Ref. \cite{b15}	&  & & &  &  &  & 1.79  & 4.97  & 1.44  & 0.074\\
													Ref. \cite{b16}	& 1.45  & 1.62 &  1.74 & 1.83 & 1.90 & 1.97  &  &  &  1.44  &  0.073   \\
													\noalign{\smallskip}\hline\noalign{\smallskip} \\
													$^{162}$Gd& & & & & & &  & & & \\
													This work & 1.45 & 1.62& 1.76 & 1.87 & 1.98 & 2.11 & 1.42 & 2.53 &  1.44 & 0.073\\
													Ref. \cite{b15} &  & &  &  &  &  &  1.76  &  4.80   & 1.44   & 0.074 \\
													Ref. \cite{b16}	& 1.45 &  1.64 & 1.77 & 1.87 & 1.96 & 2.04  &  &  & 1.44  & 0.074\\
													\noalign{\smallskip}\hline\noalign{\smallskip} \\
													$^{160}$Dy & 1.46(7) & 1.23(7)  & 1.70(16) & 1.69(9) & & & & 2.52(44) & & \\
													This work & 1.46 & 1.67 & 1.84 & 2.03 & 2.21 & 2.40 & 1.42 & 2.48 & 1.46 & 0.076\\
													Ref. \cite{b15} &  & &  &  &  &  &  1.89   &  5.70     &  1.45 & 0.075 \\
													\noalign{\smallskip}\hline\noalign{\smallskip} \\
													$^{162}$Dy & 1.42(6) & 1.48(9) & 1.70(9) & 1.72(11) & 1.62(20) & 1.62(20) & & & 1.78(16) &  0.137(12)\\
													This work& 1.45 & 1.63 & 1.76 & 1.88 & 2.00 & 2.12 & 1.42 & 2.52 & 1.44 & 0.073\\
													Ref. \cite{b15} &  & &  &  &  &  &  1.75   &   4.70    & 1.44   & 0.073 \\
													Ref. \cite{b16} & 1.45  & 1.64  & 1.77  &  1.87  & 1.96  & 2.04   &    &     & 1.44  & 0.074 \\
													\noalign{\smallskip}\hline\noalign{\smallskip} \\
													$^{164}$Dy & 1.30(7) &  1.56(7) & 1.48(9) & 1.69(9) & & & & & 2.00(27) & 0.240(33) \\
													This work& 1.44 & 1.60 & 1.73 & 1.84 & 1.95 & 2.07 & 1.43 & 2.53 &  1.45 & 0.073\\	
													Ref. \cite{b15} &  & &  &  &  &  &  1.73   &   4.55   & 1.44   & 0.073 \\
													\noalign{\smallskip}\hline\noalign{\smallskip} \\  
													Rigid rotor & 1.43& 1.57& 1.65& 1.69& 1.72& 1.74 & 1.43 & 2.57 &  1.43& 0.073\\
													\noalign{\smallskip}\hline
												\end{tabular}
											\end{center}
										\end{table}				
										
										\setlength{\tabcolsep}{4.3pt}
										\begin{table}[H]
											\caption{ The comparison of the theoretical predictions of B(E2) transition probabilities with available experimental data\cite{b33}  (upper line), for $^{166,168}$Yb, $^{228,230}$Th  and   $^{166-170}$Er isotopes . $\Delta K=0$ transition rates are normalized to the $2_{g}^{+}\rightarrow0^{+}_{g}$ transition, while $\Delta K=2$ transitions to the $2_{\gamma}^{+}\rightarrow0^{+}_{g}$ transition, as in \cite{b26,b28}.}
											\label{Tab7}
											\begin{center}
												\begin{tabular}{|c|llllllll|}
													\noalign{\smallskip}\hline\noalign{\smallskip}
													Nucleus&$\frac{4^{+}_{g}\rightarrow2^{+}_{g}}{2^{+}_{g}\rightarrow0^{+}_{g}}$& $\frac{6^{+}_{g}\rightarrow4^{+}_{g}}{2^{+}_{g}\rightarrow0^{+}_{g}}$& $\frac{8^{+}_{g}\rightarrow6^{+}_{g}}{2^{+}_{g}\rightarrow0^{+}_{g}}$& $\frac{10^{+}_{g}\rightarrow8^{+}_{g}}{2^{+}_{g}\rightarrow0^{+}_{g}}$& $\frac{12^{+}_{g}\rightarrow10^{+}_{g}}{2^{+}_{g}\rightarrow0^{+}_{g}}$&
													$\frac{14^{+}_{g}\rightarrow12^{+}_{g}}{2^{+}_{g}\rightarrow0^{+}_{g}}$& $\frac{2^{+}_{\gamma}\rightarrow2^{+}_{g}}{2^{+}_{\gamma}\rightarrow0^{+}_{g}}$& $\frac{2^{+}_{\gamma}\rightarrow4^{+}_{g}}{2^{+}_{\gamma}\rightarrow0^{+}_{g}}$\\
													\noalign{\smallskip}\hline\noalign{\smallskip}
													$^{166}$Er & 	1.45(12) & 1.62(22)  &  1.71(25) & 1.73(23) & & & & \\	 
													This work& 1.45 & 1.65 & 1.80 & 1.95 & 2.10 & 2.26 & 1.45 & 0.074\\	 
													\noalign{\smallskip}\hline\noalign{\smallskip} \\
													$^{168}$Er & 1.54(7) &  2.13(16) & 1.69(11) & 1.46(11) & & & & \\
													This work& 1.45 & 1.65 & 1.79 & 1.93 & 2.07 & 2.21 & 1.44 & 0.074\\  
													\noalign{\smallskip}\hline\noalign{\smallskip} \\
													$^{170}$Er &  &   & 	1.78(15) & 1.54(11) & & & & \\
													This work& 1.46 & 1.66 & 1.81 & 1.95 & 2.09 & 2.23 & 1.45 & 0.074\\
													\noalign{\smallskip}\hline\noalign{\smallskip} \\		
													$^{166}$Yb& 1.43(9) & 1.53(10) & 1.70(18) & 1.61(80)  & & & & \\ 
													This work& 1.49 & 1.75& 1.99 & 2.25 & 2.51 & 2.78 & 1.47 & 0.078\\
													\noalign{\smallskip}\hline\noalign{\smallskip} \\
													$^{168}$Yb &  &  &  &  &  & & & \\
													This work& 1.47 & 1.69 & 1.88 & 2.07 & 2.28 & 2.49 & 1.46 & 0.076 \\   
													\noalign{\smallskip}\hline\noalign{\smallskip} \\
													$^{228}$Th &  & & & & & & & \\
													This work& 1.46 & 1.69 & 1.89 & 2.05 & 2.24  & 2.44 & 1.47 & 0.077 \\
													\noalign{\smallskip}\hline\noalign{\smallskip} \\
													$^{230}$Th & 1.36(8) & & & & & & & \\
													This work& 1.47 & 1.70 & 1.89 & 2.09  & 2.29 & 2.50 & 1.46 & 0.077 \\
													\noalign{\smallskip}\hline\noalign{\smallskip} \\
													Rigid rotor & 1.43& 1.57& 1.65& 1.69& 1.72& 1.74& 1.43& 0.071\\
													\noalign{\smallskip}\hline
												\end{tabular}
											\end{center}
										\end{table}				
\section{Conclusion}
\label{s7}
In this work, we have investigated  nuclear shape phase transition within a conjonction between the prolate $\gamma$-rigid and $\gamma$-stable collective behaviours within the Bohr Hamiltonian with deformation dependent mass term. The deformation-dependent mass is applied simultaneously to $\gamma$-rigid and $\gamma$-stable parts of this Hamiltonian. An analytical formula for the energy spectrum of this problem, under  the Davidson potential in $\beta$ collective shape variable   and the  harmonic oscillator potential in  stiff $\gamma$-oscillations, has been derived  by making use of the asymptotic iteration method. The combined effect of the deformation-dependent mass and rigidity as well as harmonic oscillator stiffness parameters on the energy spectrum and wave function is duly analyzed. Also, electric quadrupole transition ratios and energy sprectrum of some $\gamma$-stable and prolate nuclei are calculated and compared with the experimental data as well as with other theoretical models. Predictions on the $\gamma$-rigidity of some nuclei are given. In addition, we have shown that our model has well improved the predictions in comparison with those of \cite{b15,b16}.

\section*{Appendix A: The Asymptotic Iteration Method }
\label{s8}
In this Appendix, we present basic concepts of the AIM; for more details, we refer the reader to
Refs. \cite{b17,b18}. The AIM has been proposed to solve homogeneous linear second-order differential
equations of the form
\begin{equation}
	\frac{d^2y_k(x)}{dx^2}=\lambda_0(x)\frac{dy_k(x)}{dx}+s_0(x)y_k(x),\ \lambda_0(x)\neq0
	\label{EE1}
\end{equation}
where the variables  $s_0(x)$ and  $\lambda_0(x)$ are sufficiently differentiable.
The differential equation (\ref{EE1}) has a general solutions ,
\begin{equation}
	y(x)=\exp\left(-\int^x\alpha(z)dz\right)\left[C_2+C_1\int^x\exp\left(\int^{z}\left(\lambda_0(t)+2\alpha(t)\right)dt\right)dz\right] \nonumber
	\label{EE2}
\end{equation}
where $C_1$ and $C_2$ are two constants.
If we have $k>1$, then for sufficiently large $k$  $\alpha(x)$ values can be obtained \cite{b17,b18},
\begin{equation}
	\frac{s_k(x)}{\lambda_k(x)}=\frac{s_{k-1}(x)}{\lambda_{k-1}(x)}:=\alpha(x)
	\label{EE3}
\end{equation}
with the sequences
\begin{equation}
	\lambda_{k}(x)=\frac{d\lambda_{k-1}(x)}{dx}+s_{k-1}(x)+\lambda_0(x)\lambda_{k-1}(x)
	,\ s_{k}(x)=\frac{ds_{k-1}(x)}{dx}+s_0(x)\lambda_{k-1}(x)
	\label{EE4}
\end{equation}
and the energy eigenvalues are then computed by means of the following quantization condition \cite{b17}:
\begin{equation}
	\Delta_k(x)=
	\left|
	\begin{array}{lr}
		\lambda_k(x)&s_k(x) \\
		\lambda_{k-1}(x)&s_{k-1}(x)
	\end{array}
	\right|=0\ \ ,\ \  \ k=1,2,3,\cdots
	\label{EE5}
\end{equation}
For a given potential, the procedure consists first to convert the Schr\"{o}dinger equation into the form of equation (\ref{EE1}). Then $s_0(x)$ and $\lambda_0(x)$ are determined, while $s_n(x)$ and $\lambda_n(x)$ are calculated via the recurrence relations given by equation (\ref{EE4}). The energy eigenvalues are then obtained by imposing the quantization condition shown in equation (\ref{EE5}) .
\section*{Appendix B: Calculation of the normalization constants}
\label{s9}
In this Appendix we present the  expressions of the  normalization constants for  the radial wave function.
The radial wave functions $\xi(\beta)$ are normalized through
\begin{equation}
\int_{0}^{\infty}\beta^{4-2\chi}|\xi(\beta)|^2d\beta=1
\label{EEE1}
\end{equation}
with
\begin{equation}
	\xi(\beta)=N_{n_{\beta}}\cdot\beta^{\omega_1}\left(1+a\beta^2\right)^{\omega_2} {}_2F_1\left(-n_{\beta},-n_{\beta}-\mu;-\mu-\eta-2n_{\beta};1+a\beta^2\right)
	\label{EEE2}
\end{equation}
By introducing  a new variable $t=\frac{-1+a\beta^2}{1+a\beta^2}$ and using  the relation between hypergeometrical functions and the generalized Jacobi polynomials\cite{b34,b35}, the normalization condition (\ref{EEE1}) reduces to
\begin{equation}
N_{n_{\beta}}^2 C_1 \int_{-1}^{1}a^{\frac{3}{2}-\eta}\left(1-t\right)^{2n_{\beta}+2\mu}\left(1+t\right)^{\eta}\left[P_{n_{\beta}}^{(-(1+n_{\beta}+\mu+\eta),\eta)}\left(\frac{t+3}{t+1}\right)\right]^2dt=1
\label{EEE3}
\end{equation}
This leads to
\begin{equation}
N_{n_{\beta}}^2 C_1 \int_{-1}^{1}2^{2n_{\beta}}a^{\frac{3}{2}-\eta}\left(1-t\right)^{2\mu}\left(1+t\right)^{\eta}\left[P_{n_{\beta}}^{(\mu,\eta)}\left(t\right)\right]^2dt=1
\label{EEE4}
\end{equation}
with
\begin{equation}
 C_1=\frac{2^{-2\left(n_{\beta}+1+\mu+\frac{1}{2}\eta\right)}\Gamma\left(-2n_{\beta}-\mu-\eta\right)^2\Gamma\left(n_{\beta}\right)^2 n_{\beta}^2}{a^{\frac{5}{2}}\Gamma\left(-n_{\beta}-\mu-\eta\right)^2}
 \label{EEE5}
\end{equation}
Using the following usual orthogonality relation of Jacobi polynomials \cite{b34},
\begin{equation}
\int_{-1}^{1}\left(1-t\right)^{2\alpha}\left(1+t\right)^{\beta}\left[P_{n}^{(\alpha,\beta)}\left(t\right)\right]^2dt=\frac{2^{4\alpha+\beta+1}\Gamma\left(\alpha+\frac{1}{2}\right)\Gamma\left(n+\alpha+1\right)^2\Gamma\left(2n+1+\beta\right)}{\sqrt{\pi}n!^2\Gamma\left(1+\alpha\right)\Gamma\left(2\alpha+\beta+2n+2\right)}
\label{EEE6}
\end{equation}
we finally obtain the generalized formula of the normalization constant $N_{n_{\beta}}$ :
\begin{equation}
N_{n_{\beta}}=\frac{\Gamma\left(-n_{\beta}-\mu-\eta\right)^2\sqrt{\pi}\Gamma\left(\mu+1\right)\Gamma\left(2\mu+\eta+2n_{\beta}+2\right)}{2^{2\mu-1}a^{-1-\eta}\Gamma\left(-2n_{\beta}-\mu-\eta\right)^2\Gamma\left(\mu+\frac{1}{2}\right)\Gamma\left(n_{\beta}+1+\mu\right)^2\Gamma\left(2n_{\beta}+1+\eta\right)}
\label{EEE7}
\end{equation}				
					
		\end{document}